\documentclass[pra,twocolumn,showpacs]{revtex4}
\usepackage{graphicx}
\usepackage{epstopdf}

\usepackage{amsmath}
\usepackage{amssymb}
\usepackage[T1]{fontenc}
\usepackage{calligra}
\usepackage{mathrsfs}
\newcommand{\n}{\nonumber}
\newcommand{\bn}{\begin{eqnarray}}
\newcommand{\en}{\end{eqnarray}}
\newcommand{\eml}{\end{multline}}
\newcommand{\bml}{\begin{multline}}

\newcommand{\textfrc}[1]{{\frcseries#1}}
\newcommand{\mathfrc}[1]{\text{\textfrc{#1}}}
\DeclareMathAlphabet{\mathpzc}{OT1}{pzc}{m}{it}
\DeclareFontFamily{T1}{calligra}{}
\DeclareRobustCommand\calligra{%
\renewcommand\ss{\symbol{255}\kern-.22em}%
\fontfamily{calligra}%
\fontencoding{T1}%
\selectfont}
\newenvironment{frcseries}{\fontfamily{calligra}\selectfont}{}
\begin{document}

\title {Scattering by an oscillating barrier: quantum, classical, and semiclassical comparison}

\author{Tommy A. Byrd$^{1}$, Megan K. Ivory$^{1}$, Andrew J. Pyle$^{2}$, Seth Aubin$^{1}$, Kevin A. Mitchell$^{3}$, John B. Delos$^{1}$, and Kunal K. Das$^{2}$}

\affiliation{$^{1}$ Department of Physics, College of William and Mary, Williamsburg, VA 23187, USA }
\affiliation{$^{2}$ Department of Physical Sciences, Kutztown University of Pennsylvania, Kutztown, PA 19530, USA }
\affiliation{$^{3}$ School of Natural Sciences, University of California, Merced, CA 95344, USA}

\date{\today }
\begin{abstract}
We present a detailed study of scattering by an amplitude-modulated potential barrier using three distinct physical frameworks: quantum, classical, and semiclassical. Classical physics gives bounds on the energy and momentum of the scattered particle, while also providing the foundation for semiclassical theory. We use the semiclassical approach to selectively add quantum-mechanical effects such as interference and diffraction. We find good agreement between the quantum and semiclassical momentum distributions. Our methods and results can be used to understand quantum and classical aspects of transport mechanisms involving time-varying potentials, such as quantum pumping.
\end{abstract}
 \maketitle

\date{\today }

\section{Introduction}

	Scattering dynamics involving periodic time-varying potentials is of fundamental importance to quantum transport physics and related applications in mesoscopic condensed matter physics. The quantum-mechanical treatment of an oscillating barrier was first studied by B{\"u}ttiker and Landauer in order to understand electron tunneling times \cite{ButtikerLandauer_PRL1982}, and their work built on previous work on photon-assisted tunneling in superconducting diode junctions \cite{TienGordon_PhysRev1963}. Since then several workers have developed theoretical tools for treating time-varying barrier or well potentials, for studying photon-assisted tunneling \cite{Pimpale_JPhysA1991,FedirkoVyurkov_PhysicaStatusSolidiB2000}, quantum pumping \cite{GarttnerSchmelcher_PRE2010}, and electron scattering by intense laser-driven potentials \cite{EmmanouilidouReichl_PRA2002}. These systems can display rich quantum and classical dynamics that include chaotic scattering and chaos-assisted tunneling \cite{HenselerRichter_PRE2001,linballentine,pecoralee,Steck01,Steck02}, dynamical localization \cite{GrossmannHanggi_PRL1991}, and quantum interference \cite{RahavBrouwer_PRB2006}.

Scattering by an amplitude-modulated potential barrier is of fundamental interest on its own, and it is also a building block for the more complex time-dependent potentials used in quantum pumping \cite{thouless,brouwer-1,das-PRL}. For example, the turnstile pump employs two potential barriers whose amplitudes oscillate $\pi/2$ out of phase from each other. Despite its technological promise of generating highly controlled and reversible currents at the single electron level \cite{Ferry-Goodnick}, quantum pumping in normal mesoscopic conductors remains elusive \cite{Switkes_Science1999,Brouwer_PRB2001} (though it has been recently observed in a hybrid superconducting system \cite{josehpson-quantum-pump2}).

Experimental systems based on ultracold atoms offer the possibility of conducting precision tests of quantum pumping theories, while avoiding the capacitive coupling and rectification effects that have plagued attempted solid state implementations \cite{Brouwer_PRB2001}. Furthermore, the use of ultracold atomic gases allows control over the momenta of the pumped particles and the coherence of the gas, and it permits precision imaging and velocity measurements, as well as the choice between Bose-Einstein and Fermi-Dirac statistics.

In this paper, we study classical, semiclassical, and quantum dynamics of one-dimensional scattering by an amplitude-modulated Gaussian barrier. Motivated by possible experimental implementations with ultracold atoms, our main theoretical results are based on calculations of the scattered momentum distribution for atomic wavepackets of well-defined incident velocity, such as propagating Bose-Einstein condensates (BEC). By employing a semiclassical formalism, we start with the classical dynamics and selectively turn on quantum processes such as interference and diffraction. Our main results can be summarized as follows: i) Classical physics gives bounds on the range of scattered momentum states; ii) Semiclassical and full quantum calculations predict similar final momentum distributions; (iii) The heights of Floquet peaks, which are not easily predicted by quantum calculations, are explained quantitatively by the semiclassical method. Interestingly, the physical pictures for the scattering process are quite different for the semiclassical and quantum methods. The semiclassical approach interprets the discrete final momentum values as inter-cycle interference over multiple barrier oscillations, but with the relative amplitudes of these states determined by intra-cycle interference. In contrast, from the Floquet perspective of full quantum theory, the final momentum states can be viewed as sidebands of the initial momentum state.

The paper is structured as follows: We present our model in Sec.~\ref{sec:model}, and in Sec.~\ref{sec:quantum_classical} display results of quantum and classical calculations for this model. Sec.~\ref{sec:semiclassical_description} explains the algorithm used for the semiclassical calculation, and Sec.~\ref{sec:discussion} compares and discusses the semiclassical and full quantum methods. In Sec.~\ref{sec:experiment_section}, we show how the model and results of this paper can be tested experimentally with ultracold atoms. Sec.~\ref{sec:conclusion_section} summarizes our main results. Appendices~\ref{sec:semiclassical_theory} and~\ref{sec:kevin} fill in the details of the semiclassical algorithm, and Appendix~\ref{sec:bounds} explains the range of scattered momenta based on a simpler potential.

\section{Model}\label{sec:model}
Our model is motivated by recent proposals \cite{Das2011, DasAubin_PRL2009} to simulate mesoscopic transport processes by studying ultracold atomic wavepackets propagating in quasi-one-dimensional waveguides that scatter from well-defined, localized potentials. A laser beam, blue-detuned from an atomic resonance, and tightly focused at the center of the wave guide, can create a potential barrier with a Gaussian profile, its width determined by the
laser spot size and its amplitude by the intensity of the laser.

We choose a 1D Gaussian barrier, centered at the origin, whose amplitude is modulated sinusoidally at frequency $\omega$, with potential energy $U(x,t)$ given by
\bn U(x,t)=U_0(1+A\sin(\omega t+\phi))e^{-x^2/(2\sigma^2)}\label{eqpotential}.\en
$U_0$ is the average amplitude of the barrier, $A$ is the relative modulation amplitude, $\sigma$ is the standard deviation width of the barrier, and $\phi$ is the phase of the modulation. The Hamiltonian describing particle motion and scattering from this potential is
\begin{align}
H=\frac{p^2}{2m}+U\left(x,t\right).\label{Hameq}
\end{align}

We use wavepackets with initial momentum $p_0>0$, centered at a point $\bar{x}$ far to the left of the barrier, and whose position-space wavefunction is given by
\begin{align}
\Psi\left(x,t=0\right) = F\left(x\right)e^{ip_{0}x}
\end{align}
where $F(x)$ is the envelope of the wavepacket and is typically a Gaussian of width $\beta$,
\begin{align}
F\left(x\right) = F_{G}\left(x\right) = \frac{1}{\left(2\pi\right)^{1/4}}e^{-\left(x-\bar{x}\right)^{2}/4\beta^{2}}\label{initgauss}.
\end{align}
Alternatively, the envelope may have a Thomas-Fermi distribution of radius $\beta$, such that
\begin{align}
F(x)=F_{TF}\left(x\right)= \left\{
     \begin{array}{cc}
       \sqrt{\beta^{2} - \left(x-\bar{x}\right)^{2}} &, \left|x - \bar{x}\right| < \beta\\
       0 &,\left|x - \bar{x}\right| > \beta
     \end{array}
   \right.\label{inittf}
\end{align}
The Thomas-Fermi and Gaussian envelopes are typical of BEC wavefunctions in strongly interacting and non-interacting limits, respectively. Unless otherwise noted, we employ wavepackets that are much wider than the barrier width ($\beta \gg \sigma$), with packet width $\beta$ sufficiently large such that $\beta \gg 2\pi p_{0}/m \omega$, ensuring that many barrier oscillations occur while the packet interacts with the barrier.

In the rest of the paper, unless otherwise mentioned, we use $U_{0} = m = \hbar = 1$, $A = 0.5$, $\sigma = 10$, and $\beta = 300$. The values of the incident momentum are in the range $p_{0} \simeq 1-2$, the oscillation frequency $\omega \simeq 0 - 0.2$, and in most cases the phase, $\phi$, is set equal to $0$. In the case of a Gaussian packet, we select $\bar{x} = -1500$ to ensure separation of the initial packet from the barrier.

The choice of a theoretical unit convention based on $\hbar=1$ and $m=1$ is equivalent to selecting an arbitrary time unit $t_{u}$ and a related length unit $l_{u}=\sqrt{\hbar t_{u}/m}$,  with $\hbar=1.054 \times10^{-34}$ J$\cdot$s. The corresponding energy unit is $E_{u}=\hbar/t_{u}$, while the mass unit is that of the particle, $m_{u}=m$.

\section{Quantum and Classical Calculations}\label{sec:quantum_classical}

\subsubsection{Quantum Description}\label{subsubsec:quantum_description}

\begin{figure}[t]
\includegraphics*[width=\columnwidth]{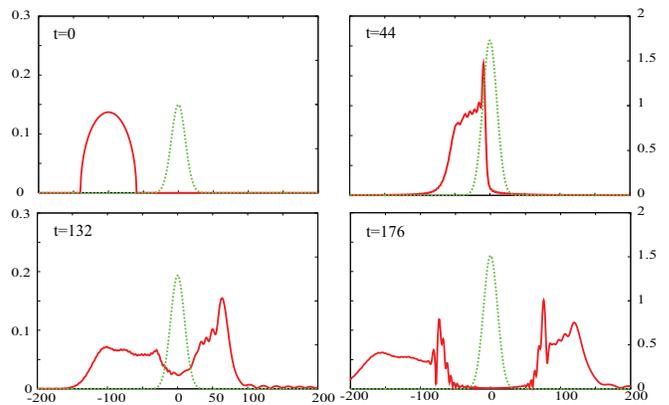}
\caption{(Color Online) Snapshots from a typical quantum-mechanical calculation
showing a Thomas-Fermi (Eq. \eqref{inittf}) wavepacket (left axis; solid red line) scattering off
a Gaussian barrier (right axis; dotted green line). The amplitude of the barrier varies in
time according to Eq. \eqref{Hameq} with $U_0=1, A=0.9$. The barrier width ($\sigma=10$) here is typical in our simulations, but the packet width ($\beta=40$) is much less (to show more details) than used elsewhere ($\beta=300$) in the paper.}\label{Figure-1}
\end{figure}

We consider both quantum-mechanical and classical descriptions of the scattering process. This dual framework allows us to distinguish the classical and quantum nature of a variety of scattering features.

Our quantum-mechanical approach is based on propagating the wavepacket with the Schr\"odinger equation:
\begin{align}
-\frac{\hbar^{2}}{2m}\partial^{2}_{x}\Psi +U\left(x,t\right)\Psi = -i\hbar \partial_{t}\Psi
\end{align}
via a split-step operator method \cite{G.P.Agarwal} that incorporates
the time-variation of the scattering potential $U(x,t)$.  The numerical calculation is done using a Fast-Fourier Transform (FFT) in a parallelized routine in FORTRAN. With periodic boundary conditions implicit in the FFT, the spatial range $R$ (typically $\sim 8000$ in dimensionless units) is chosen sufficiently large to allow the entire wavepacket to interact with the barrier at $R/2$ without significant wraparound. The spatial grid density and the time step for propagation are both taken to be of the order of $~0.1$ in dimensionless units.  The resulting momentum grid density $2\pi/R\simeq 10^{-3}$ is more than sufficient to resolve the narrowest momentum space features that we encounter.

Figure~\ref{Figure-1} shows a quantum calculation of a Thomas-Fermi wavepacket in position
space at four separate times as it scatters from an
amplitude-modulated Gaussian barrier. In order to show more details of the scattering, the packet width shown in this figure is intentionally more narrow than that used in the rest of the paper.  The resulting transmitted and reflected wavepackets show considerable structure, but with no clear pattern, except for some residual spatial oscillation suggesting some type of interference effect. While examining the scattering process in position space does not yield any simple clues regarding its dynamics,
the momentum-space picture offers significantly more insight into the relevant physics.

To obtain the wavefunction in momentum space, at a chosen large time, $t = t_{f}$, after the packet has moved away from the potential barrier, we compute the Fourier transform of $\Psi(x,t_f)$ :
\begin{align}
\tilde{\Psi}\left(p,t_f\right) = \frac{1}{\sqrt{2\pi}} \int^{\infty}_{-\infty} e^{-ipx} \Psi\left(x,t_f\right)dx.
\end{align}
We also compute the corresponding final-momentum probability density,
\begin{align}
\tilde{P}_{Q}^{F}\left(p_f\right) = |\tilde{\Psi}(p_f,t_{f})|^{2}.\label{pqtilde}
\end{align}
Here, $p_f$ is used to indicate momentum at the chosen final time. Also, we note that for sufficiently large times, such that the packet has moved far from the barrier, the final momentum distribution is constant in time, while the momentum-space wavefunction is not.

A time-periodic potential produces energy and momentum sidebands to the incident carrier momentum state, which can be described by Floquet theory, the temporal analog of Bloch's theorem. In our model, a wavepacket is incident on the barrier with fixed group momentum $p_{0}$ and associated kinetic energy $E_{0} = p_{0}^{2}/(2m)$. Since we use spatially-broad packets, the incident packet has a very narrow momentum spread. The interaction of the incident wavepacket with the amplitude-modulated barrier produces a series of discrete momentum states separated in energy by $\hbar\omega$. The allowed final-momentum states must obey the equation
\begin{align}
p_{f}(n) = \pm\sqrt{2m\left(E_{0}+ n\hbar \omega\right)}\label{pfinalsanalytic}
\end{align}
where $n$ is any integer satisfying $n\geq -E_{0}/\hbar \omega$, and with $(+)$ and $(-)$ corresponding to transmission and reflection, respectively.

\begin{figure}[t]
\includegraphics*[width=\columnwidth]{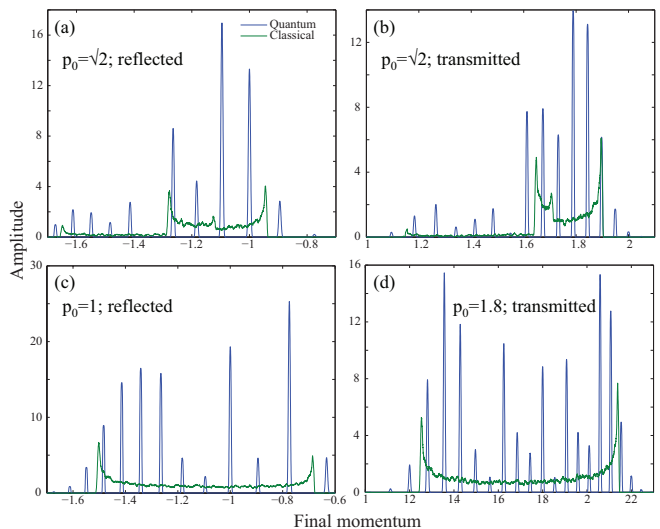}
\caption{(Color Online) Quantum (sharply peaked curves, blue online) and classical (green online) momentum distribution for fixed $\omega=0.1$, $U_0=1$, $A=0.5$, $\sigma=10$, $\beta=300$, but different incident packet velocities, $p_0$. The classical distributions were obtained via the histogram method, and statistics account for the fluctuations seen in the curves. (a) The reflected and (b) transmitted parts for $p_0=1.4142$; (c) reflected part for $p=1.0$, when transmission is negligible; (d) transmitted part $p_0=1.8$, when reflection is negligible.}\label{Figure-2}
\end{figure}

Fig.~\ref{Figure-2} shows the momentum-space distribution of the reflected and transmitted wavepackets after scattering from the amplitude-modulated barrier. The results of the full quantum calculation show the regular "comb" of discrete momentum states consistent with Eq. \eqref{pfinalsanalytic}. Fig.~\ref{Figure-2} also plots the classical momentum-space distribution for a Gaussian ensemble of particles with the same initial momentum spread as the initial quantum wavepacket (see next subsection for details). The classically-allowed bounds for the final momentum roughly constrain the Floquet comb on both reflection and transmission, though we find that the comb often extends slightly past the classically-allowed bounds. However, the amplitude of the teeth of the comb do not appear to have any obvious pattern, and only loosely follow the strength of the classical final-momentum distribution.

 The semiclassical approach presented in Sec.~\ref{sec:semiclassical_description} and the Appendices will provide an alternative explanation for the positions of the teeth of the Floquet comb in terms of inter-cycle interference, and will provide an explanation for the relative amplitudes of the comb teeth in terms of intra-cycle interference.

\subsubsection{Classical Description}\label{subsubsec:classical_description}

The classical description of the scattering dynamics computes trajectories based on the Hamiltonian of Eq. \eqref{Hameq}. In the static limit, particles of incident energy above $U_0$ are transmitted, and those below are reflected. In contrast, scattering from an oscillating barrier leads to significant changes in the particle momentum distribution, as particles gain or lose energy with the rise and fall of the potential. The final outcome depends on the phase of the oscillation as the particle encounters the barrier, and must generally be computed numerically.

\begin{figure}[t]
\includegraphics*[width=\columnwidth]{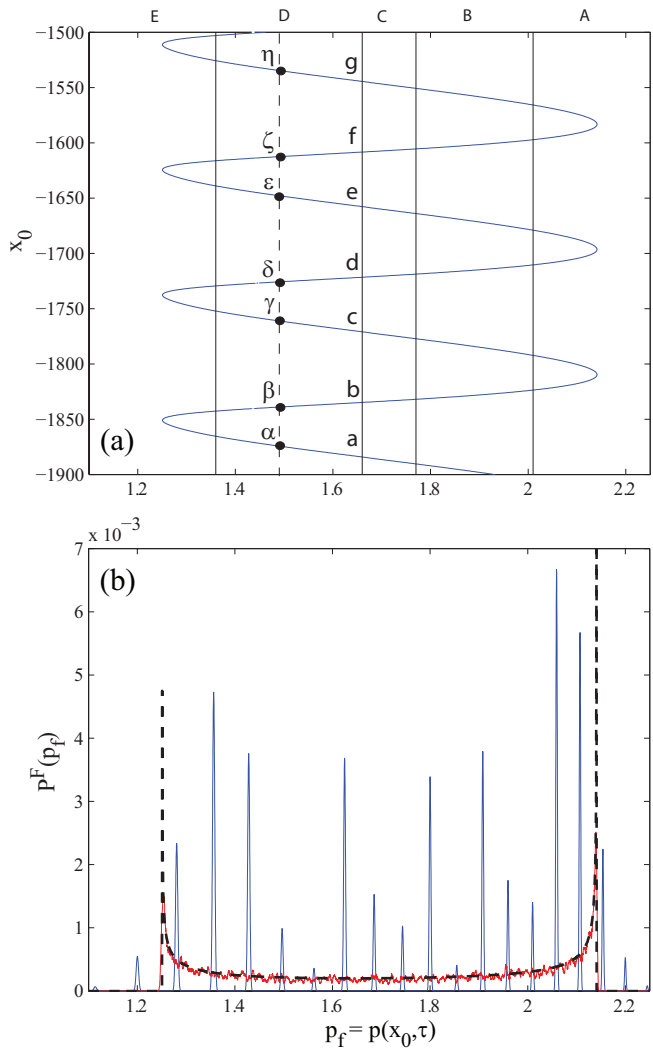}
\caption{(Color Online) (a) Final momentum vs. initial position for $A = 0.5$, $\omega = 0.1$, $p_{0} = 1.8$. Capital letters correspond to different momentum regions (separated by solid vertical lines; see App.~\ref{sec:semiclassical_theory}). At a selected $p_f$, marked by the dashed line, paths arrive after beginning at many different $x_0$;  those points are labeled by Greek letters.  Each lies on a branch of the multivalued function $x_0(p_f,t_f)$, and each branch is labeled by a Roman letter. (b) Final-momentum distributions calculated quantum-mechanically and classically. The classical calculations show a histogram (solid line, red online) and $P_{C}^{F}(p_{f})$ from Eq. \eqref{pctilde} (dashed curve, black online). Fluctuations in the histogram arise for statistical reasons.}\label{fig3regions}
\end{figure}

Our quantum and semiclassical calculations diminish the role of the phase of the barrier oscillation by studying Heisenberg-limited wavepackets with a large position spread and a well defined momentum, so that many barrier oscillations occur while the wavepacket is interacting with it. We mimic such wavepackets in our classical approach by employing ensembles of particles with initial conditions whose position and momentum distributions, $P^0_C\left(x\right)$ and $\tilde{P}^0_C\left(p\right)$, match those of the quantum distributions:

\begin{subequations}
\begin{align}
P_{C}^{0}\left(x\right)=\left|\Psi\left(x,t=0\right)\right|^{2}\\
\tilde{P}_{C}^{0}\left(p\right)=\left|\tilde{\Psi}\left(p,t=0\right)\right|^{2}
\end{align}
\end{subequations}

Generally, our initial momentum distributions are sufficiently narrow that classical particles can begin with a fixed initial momentum, distributed along a line segment that substantially covers the width of the initial wave packet, with statistical weights $P_C^0(x)$.

\begin{figure}[t]
\includegraphics*[width=\columnwidth]{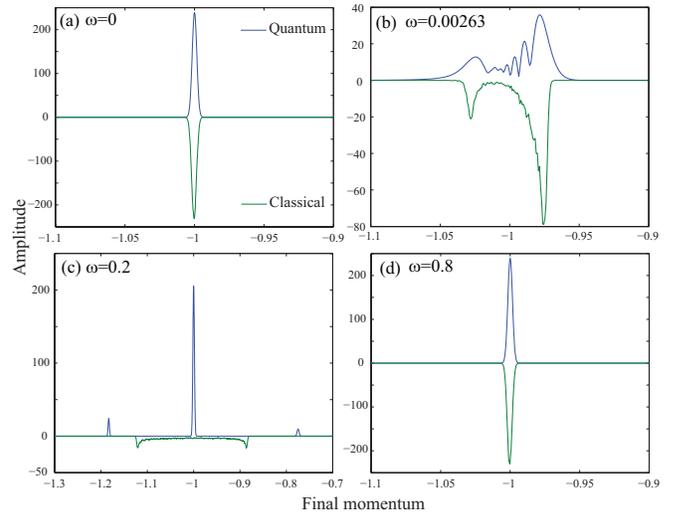}
\caption{(Color Online) Momentum distributions for fixed velocity of incident packet but for different values of $\omega$. The correlation between classical and quantum distributions reflected in Fig.~\ref{Figure-4} is seen. Comparison of quantum (blue and above axis) and classical (green and below axis) momentum distributions for $p_0=1.0$ with $\omega=0, 0.00263, 0.2, 0.8$. Quantum and classical results are correlated for low and high values of $\omega$, with significant differences appearing at intermediate values.  These correlations are quantified in Fig. \ref{Figure-4}.}\label{Figure-geez}
\end{figure}

The distribution $\tilde{P}_C^F(p_f)$ of final momenta $p_f$ can be obtained by numerically integrating trajectories and grouping them in bins of final momentum to plot a histogram, as shown in Fig.~\ref{Figure-2} and Fig.~\ref{fig3regions}(b). Alternatively, we can compute trajectories numerically to obtain the final momentum as a function of initial position $x_0$ and final time $t_f$, $p_f=p(x_0,t_f)$, as shown in Fig.~\ref{fig3regions}(a). We note that due to the periodicity of the barrier amplitude, $p_f$ is a continuous periodic function of $x_0$, with period $2\pi p_0 / \omega m$. Any such periodic function has a maximum and minimum, which define the classically-allowed range of $p_f$, as shown in Fig.~\ref{fig3regions}. Furthermore, this periodicity means that many initial positions $x^j_0(p_f,t_f)$ contribute to the final momentum distribution $\tilde{P}_C^F(p_f)$. Each $x_{0}^{j}(p_{f},t_{f})$ contributes to $\tilde{P}_{C}^{F}(p_f)$ a term proportional to $|\partial x_{0}^{j}/ \partial p_f| = \left.|\partial p(x_{0},t_{f})/\partial x_{0}|^{-1}\right|_{x_{0}=x_{0}^{j}(p_f,t_{f})}$, so
\begin{align}
\tilde{P}_{C}^{F}\left(p_f\right) = \sum_{j} P_{C}^{0}\left(x_{0}^{j}\left(p_{f},t_{f}\right)\right) \left|\partial x_{0}^{j}/\partial p_f\right|\label{pctilde}
\end{align}

Figure~\ref{fig3regions} shows the final classical momentum distribution $\tilde{P}_C^F(p_f)$ computed by both the histogram method  (solid line, red online) and according to Eq. \eqref{pctilde} (dashed curve, black online), as well as the final quantum momentum distribution. The maximum and minimum of $p_f$ define the classically-allowed region, with $\partial p_f/\partial x_0$ going to zero at these locations, and its reciprocal in Eq. \eqref{pctilde} tending to infinity \cite{sing}.

When we compare the quantum calculation to this classical calculation (Figs.~\ref{Figure-2} and~\ref{fig3regions}(b)) we see that the boundaries of the classically-allowed region accurately define the region of momentum space in which Floquet peaks are large. Small peaks also appear outside but close to the classically-allowed region. As we show in the semiclassical treatment of Sec.~\ref{sec:semiclassical_description}, these are the result of momentum-space tunneling (or diffraction) into the classically-forbidden region.

\begin{figure}[t]
\includegraphics*[width=\columnwidth]{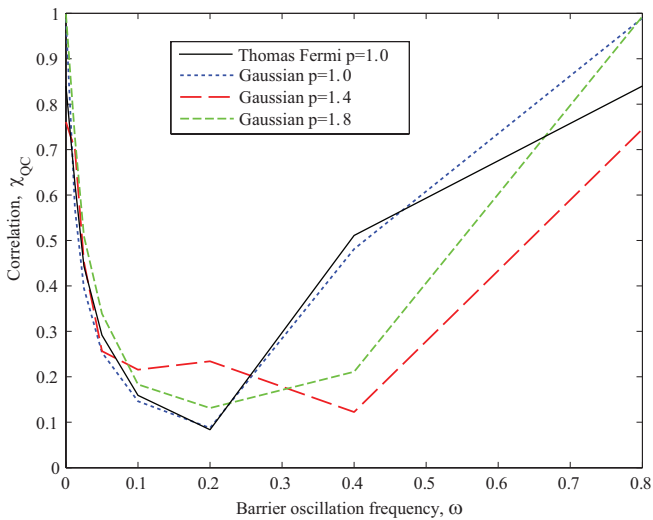}
\caption{(Color Online) Correlation coefficient $\chi_{QC}$ of the normalized classical and quantum momentum densities (defined in Eq. \eqref{correq}) plotted as a function of the barrier oscillation frequency $\omega$. The three different incident momentum represent: (i) Primarily transmitting ($p_0=1.8$), (ii) primarily reflecting ($p_0=1.0$)  and (iii) a transition regime ($p_0=1.4142$) of partial reflection and partial transmission. All use Gaussian packets, except that for $p_{0}=1.0$. The result of using a Thomas-Fermi packet is also plotted, showing that the choice of packet shape is not crucial if the packets are sufficiently broad.}\label{Figure-4}
\end{figure}

We also find that the barrier oscillation frequency $\omega$, an easily variable experimental parameter, can be used to control the concurrence of the classical and quantum calculations, with good agreement in the limits of very high and low frequencies. For a static barrier, or for extremely low frequencies, momentum conservation in classical and quantum theories ensures agreement. As the frequency is increased, keeping the initial packet unchanged, the agreement gets poorer (Figs.~\ref{Figure-geez}(b) and~\ref{fig3regions}(b)). The classical momentum distribution broadens, and the quantum distribution acquires a ``comb'' structure since Floquet peaks begin to resolve as their separations become greater than their widths (which depend inversely on the width of the initial packet in position space). This is the range of particular interest in this paper.  At very high frequencies, the incident particles cannot respond fast enough to the modulation of the barrier, and so they effectively interact with the time-average of the potential. The classically-allowed region narrows, while in the Floquet picture, the spacing between the Floquet peaks increases (Fig.~\ref{Figure-geez}(c)).  When there is only one non-negligible Floquet peak remaining, it coincides with the classically-allowed region, resulting again in good agreement between the two methods (Fig.~\ref{Figure-geez}(d)).

\begin{figure}[t]
\includegraphics*[width=\columnwidth]{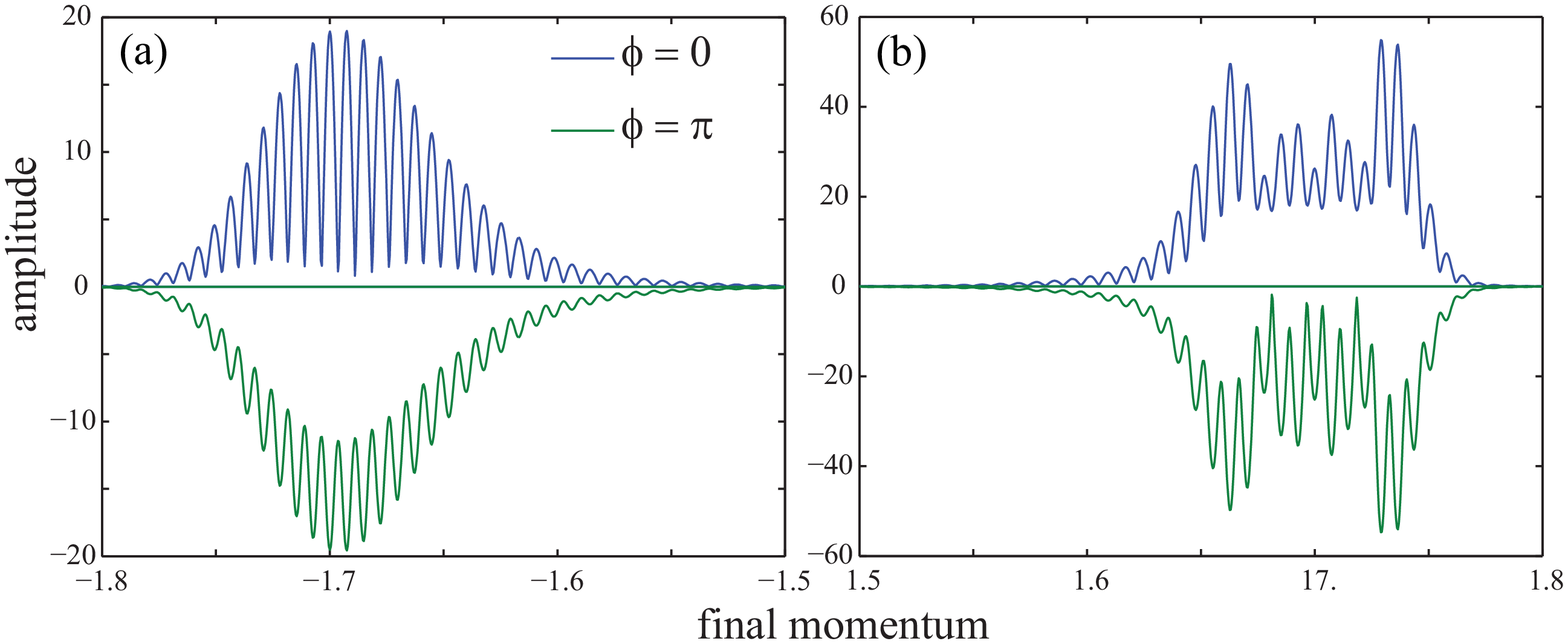}
\caption{(Color Online) Phase sensitivity for very low omega values: even for  the same $\omega=0.0125$  the momentum distribution changes with phase of the barrier oscilation: $\phi=0$ (blue, above axis) and $\phi=\pi$ (green, below axis) with (a) showing transmitted fraction and  (b) reflected fraction. The incoming wave packet was Gaussian-shaped, with $p_{0} = 1.7$ and $\beta = 300$, and the barrier parameters were $\sigma = 10$ and $A=0.5$.}\label{Figure-5}
\end{figure}

In order to quantify the comparison of the final quantum momentum distribution, Eq. \eqref{pqtilde}, with the final classical momentum distribution, Eq. \eqref{pctilde}, we define a kind of final momentum-density correlation coefficient,
\bn \chi_{QC}
&=&\frac{\int dp_{f}\tilde{P}_Q^{F}(p_{f})\tilde{P}_C^{F}(p_{f})}{\sqrt{\int dp_{f} \left[\tilde{P}_Q^{F}(p_{f})\right]^{2}\int dp_{f}\left[\tilde{P}_C^{F}(p_{f})\right]^{2}}}.\label{correq}\en

This correlation coefficient is plotted in Fig.~\ref{Figure-4}, which confirms the behavior indicated above, wherein the quantum and the classical distributions are in close agreement at low and at high frequencies, but not at intermediate frequencies.

At very low frequencies, the quantum momentum distribution depends upon the initial phase $\phi$ of the potential (see Eq. \eqref{eqpotential}), as we show in Fig.~\ref{Figure-5}. At such low frequencies, the incident packet interacts with the barrier over only a fraction of a cycle, thus experiencing a barrier amplitude that is strongly dependent on the oscillation phase.

To summarize, we see that classical calculations describe the range of momenta over which Floquet peaks are large, and they agree with quantum calculations at very high and very low frequencies, but more generally the heights of the Floquet peaks in the quantum calculations remain mysterious. They will be explained using a semiclassical method described in the next section.

\section{Semiclassical Description}\label{sec:semiclassical_description}

It is a general principle of quantum mechanics \cite{Feyn2} that when in classical mechanics we add probabilities associated with different paths leading to the same final state as in Eq. \eqref{pctilde}, in quantum mechanics we add amplitudes. In the semiclassical approach, each amplitude is the square root of the classical density combined with a phase. In the present case, Eq. \eqref{pctilde} is replaced by
\begin{align}
\tilde{P}_{SC}^{F}\left(p_{f}\right) &= \left|\tilde{\Psi}_{SC}\left(p_f,t_{f}\right)\right|^{2}\mbox{,     with}\\
\begin{split}
\tilde{\Psi}_{SC}\left(p_{f}, t_{f}\right) &= \sum_{j} F\left(x_{0}^{j}\left(p_{f},t_{f}\right)\right)
\left|\tilde{\mathcal{J}}_{j}\left(p_{f},t_{f}\right)\right|^{-1/2}\\
&\times  \exp{\left(i\left[\tilde{\mathcal{S}}_{j}\left(p_{f},t_{f}\right)/\hbar - \tilde{\mu}_{j} \pi/2\right]\right)}\label{psitildesum}
\end{split}
\end{align}
where we are again using $p_f=p\left(x_0,t_f\right)$. $F(x_{0})$ is the envelope of the initial wave packet, either $F_{G}(x_{0})$ in Eq. \eqref{initgauss} or $F_{TF}(x_{0})$ in Eq. \eqref{inittf}, and $x_{0}^{j}(p_{f},t_{f})$ has the same meaning as in the paragraph above Eq. \eqref{pctilde}: trajectories that arrive at any one $p_{f}$ began from a large number of discrete $x_{0}^{j}(p_{f},t_{f})$.

Re-examining Fig.~\ref{fig3regions}(a), and thinking about $x_{0}(p_{f},t_{f})$ as a smooth but multivalued function of $p_{f}$, we divide the points $x_{0}^{j}(p_{f},t_{f})$ into intracycle and intercycle groups, where a cycle is one period of $p_f$. In Fig.~\ref{fig3regions}(a), we may say that the pair of points $\left(\alpha,\beta\right)$ belongs to one cycle, the pair $\left(\gamma,\delta\right)$ to another cycle, etc. Alternatively, we may say that the pair $\left(\beta,\gamma\right)$ belongs to one cycle, $\left(\delta,\epsilon\right)$ to the next, etc. Summing over all the points $x_{0}^{j}\left(p_{f},t_{f}\right)$ then means summing over points on distinct branches of $x_{0}\left(p_{f},t_{f}\right)$ within a cycle, and then summing over cycles. Thus the index $j$ may become a composite index, $j = \left(b,c\right)$ where $b$ is an integer labeling a branch within a cycle, and $c$ is an integer labeling the cycle.

\begin{figure}[t]
\includegraphics*[width=\columnwidth]{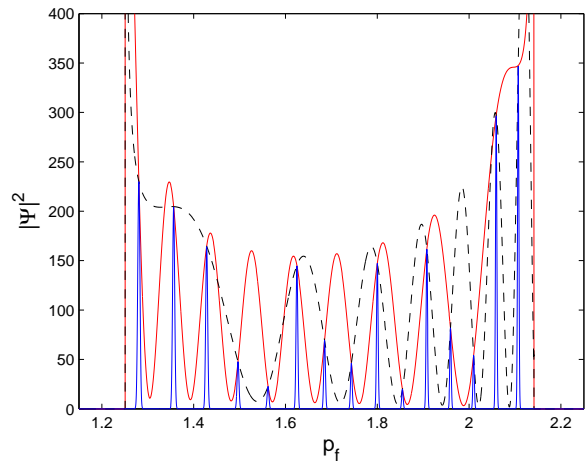}
\caption{(Color Online) Probability distributions of final momenta. The sharp peaks (blue online) are obtained by summing over all branches of all cycles. Their heights are all multiplied by the same constant so that they are comparable to the other two curves. The oscillating curves are obtained by combining two branches of a single cycle, but with different definitions of the cycle.  The solid curve (red online) corresponds to a cycle spanning branches $(b,c)$ in Fig.~\ref{fig3regions}(a), and the dashed curve (black online) is for a cycle spanning branches $(c,d)$. Where those two curves intersect, the different cycles add in phase with each other, producing the sharp peaks.}\label{singles}
\end{figure}

$\tilde{\mathcal{J}}\left(p_{f},t_{f}\right)$ is  a Jacobian, which in the present case is the same derivative defined in Eq. \eqref{pctilde},
\begin{align}
\tilde{\mathcal{J}}_{j}\left(p_{f},t_{f}\right) = \left|\frac{\partial p_{f}\left(x_{0},t_{f}\right)}{\partial x_{0}}\right|_{x_{0} = x_{0}^{j}\left(p_{f},t_{f}\right)}
\end{align}

Since $p_{f}$ is a periodic function of $x_{0}$, the values of this derivative depend on the branches within a cycle, but do not depend on which cycle is examined: $\tilde{\mathcal{J}}_{\left(b,c\right)}\left(p_{f},t_{f}\right)$ depends on the branch $b$ but is independent of the cycle $c$. In Fig~\ref{fig3regions}(a), $\tilde{\mathcal{J}}_{\alpha}\left(p_{f},t_{f}\right) = \tilde{\mathcal{J}}_{\gamma}\left(p_{f},t_{f}\right) = \tilde{\mathcal{J}}_{\epsilon}\left(p_{f},t_{f}\right) = ...$, while $\tilde{\mathcal{J}}_{\beta}\left(p_{f},t_{f}\right) = \tilde{\mathcal{J}}_{\delta}\left(p_{f},t_{f}\right) = \tilde{\mathcal{J}}_{\zeta}\left(p_{f},t_{f}\right) = ...$

$\tilde{\mathcal{S}}_{j}\left(p_{f},t_{f}\right)$ is a classical momentum-space action integrated along the path from $x_{0}^{j}\left(p_{f},t_{f}\right)$ to the final point. This integral is
\begin{align}
\tilde{\mathcal{S}}_{j}\left(p_{f},t_{f}\right) = &- \int{xdp} - \int{Edt}\n\\
\begin{split}
=&-\int_{0}^{t_{f}}{x\left(x_{0},t\right)\frac{dp\left(x_{0},t\right)}{dt}dt}\\
&-\int_{0}^{t_{f}}{E\left(t\right)dt}\label{action}
\end{split}
\end{align}

There is a simple relationship between the values of $\tilde{\mathcal{S}}_{\left(b,c\right)}\left(p_{f},t_{f}\right)$ for different cycles at fixed $p_{f}$. Let label $c$ increase with decreasing $x_{0}^{\left(b,c\right)}$; i.e., it increases by $1$ with each successive cycle of the oscillating barrier. Then
\begin{align}
\tilde{\mathcal{S}}_{\left(b,c+N\right)}\left(p_{f},t_{f}\right) = \tilde{\mathcal{S}}_{\left(b,c\right)}\left(p_{f},t_{f}\right) +N\Delta ET,\label{Scycles}
\end{align}
where $T$ is the period of one oscillation, $N$ is the number of periods separating the cycles, and $\Delta E$ is the change of energy of the particle
\begin{align}
\Delta E = \left(p_{f}^{2} - p_{0}^{2}\right)/2m.
\end{align}

Finally, we introduce the Maslov index $\mu_j$ associated with each branch of $x_{0}\left(p_{f},t_{f}\right)$. The rule for determining it is given in Appendix B. Here let it suffice to say that in Fig.~\ref{fig3regions}(a), $\tilde{\mu}_{j}$ can be taken to equal one on branches $a,c,e,g,...$ and equal to zero on branches $b,d,f,...$.

In our calculations, we compute the final momentum as a function of initial position $p_{f}\left(x_{0},t_{f}\right) = p\left(x_{0},t_{f}\right)$, then for each $p_{f}$ we identify initial points $x_{0}^{\left(b,c\right)}\left(p_{f},t_{f}\right)$ for all branches $b$ within a \emph{single} cycle $c$. For each of them we find $\tilde{\mathcal{J}}_{b}\left(p_{f},t_{f}\right)$, $\tilde{\mu}_{b}$, and $\tilde{\mathcal{S}}_{\left(b,c\right)}\left(p_{f},t_{f}\right)$ for that particular cycle. We then calculate $\tilde{\mathcal{S}}_{\left(b,c\right)}\left(p_{f},t_{f}\right)$ for other cycles using Eq. \eqref{Scycles}, and then compute the sum Eq. \eqref{psitildesum} numerically. Steps are also taken to correct the semiclassical approximation near divergent points, and the calculation is extended into the classically-forbidden regions; this procedure incorporates diffraction, or momentum space tunneling, into the semiclassical dynamics. Derivation and additional details of the semiclassical method are given in Appendix~\ref{sec:semiclassical_theory}.

Terms in the sum over cycles add with incommensurate phases, and tend to cancel unless $\Delta E = 2 \pi K$ where $K$ is any integer. This condition explains the Floquet picture introduced earlier: \emph{the momentum distribution becomes a "comb" function, with the "teeth" occuring at momenta that satisfy the commensurate phase condition,}
\begin{align}
\frac{p_{f}^2}{2m} = \frac{p_{0}^{2}}{2m} + \frac{2\pi K \hbar}{T}.\label{pstwo}
\end{align}

In Fig.~\ref{singles} we show the absolute squares of two single-cycle wavefunctions, one using branches $\left(b,c\right)$ (solid curve, red online) in Fig.~\ref{fig3regions}(a), the other using branches $\left(c,d\right)$ (dashed curve, black online). These single-cycle probabilities intersect at momenta satisfying Eqs. \eqref{pfinalsanalytic} and \eqref{pstwo}. The relative amplitudes of these intersections are determined by both the classical densities and the differences in momentum-space action, Eq. \eqref{action}, among the paths contributing to the wavefunction at each $p_f$.

\begin{figure}[t]
\includegraphics*[width=\columnwidth]{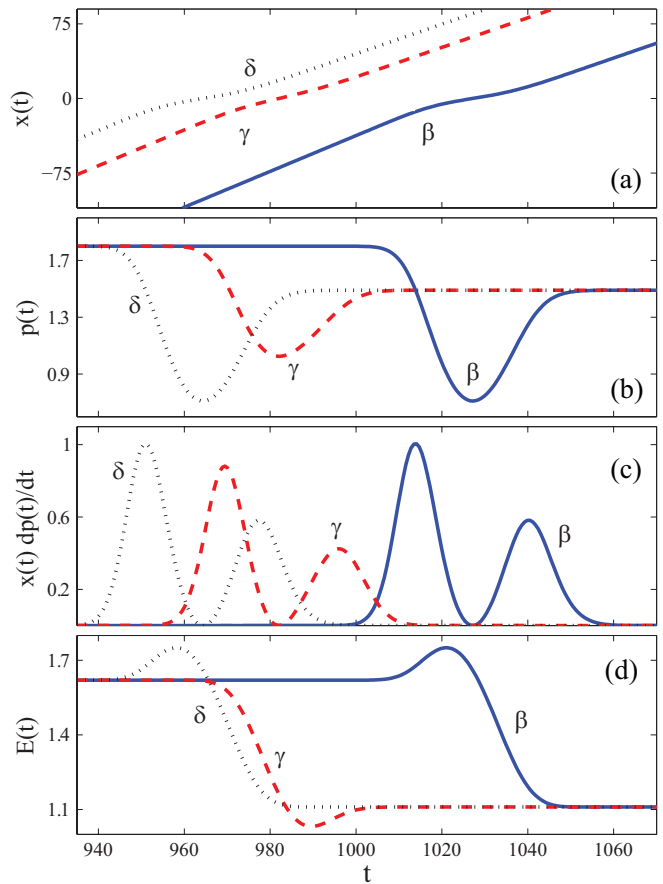}
\caption{(Color Online) Quantities that determine the phase evolution and interference of three trajectories ending with the same final momentum. The solid (blue online), dashed (red online), and dotted curves (black online) correspond to the trajectories associated with $(p_f,x_0) = \beta$, $\gamma$, and $\delta$ in Fig.~\ref{fig3regions}(a), respectively. One may think of the $(\beta,\gamma)$ trajectories as being from a single cycle, with the $\delta$ trajectory one cycle ahead of the $\beta$ trajectory. (a) Position versus time. Each trajectory shows a decrease in velocity as the barrier is initially encountered near $x=0$.  (b) Momentum versus time. (c) $x(t) dp(t)/dt$ term in the momentum-space action (Eq. \ref{action}) versus time. (d) Energy term in the momentum-space action (Eq. \eqref{action}) versus time.}\label{alphabeta}
\end{figure}

Fig.~\ref{alphabeta} shows quantities that determine the phase differences and interference for three trajectories ending with the same final momentum. Figs.~\ref{alphabeta}(a) and~\ref{alphabeta}(b) show the position and momentum, respectively, versus time. Both plots show that particles see a decrease in velocity (and momentum) as they approach the potential barrier. Figs.~\ref{alphabeta}(c) and ~\ref{alphabeta}(d) illustrate the differences in the momentum-space action, Eq. \eqref{action}. The differences in areas under the curves determine the phase differences between pairs of trajectories.  Interference associated with phase differences related to $E(t)$ for different cycles (Fig.~\ref{alphabeta}(d)) produces Floquet peaks. Phase differences between pairs of trajectories in the same cycle (Figs.~\ref{alphabeta}(c) and~\ref{alphabeta}(d)) give the interference that determines relative heights of Floquet peaks.

When we sum over cycles, the resulting probability is sharply peaked at the locations where the single-cycle probabilities intersect (Fig.~\ref{singles}), and \emph{the heights of the peaks correspond to the relative magnitude of the single-cycle probability at these locations}. Finally, we have an explanation for the relative heights of the Floquet peaks.

\section{Case Studies}\label{sec:discussion}

\begin{figure}[t]
\includegraphics*[width=\columnwidth]{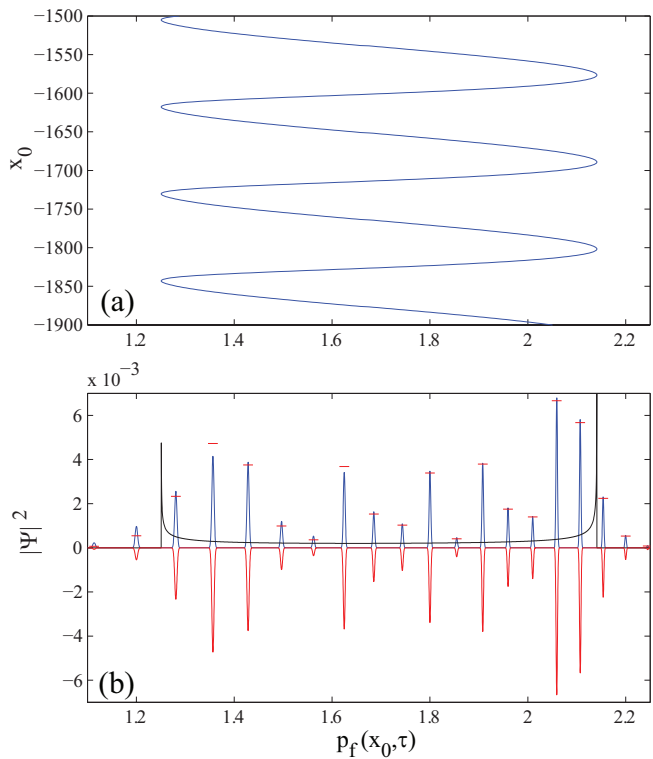}
\caption{(Color Online) (a) Final momentum vs initial position for the $\omega = 0.1$, $p_{0} = 1.8$ case. (b) Comparison of classical (plotted upwards, black online), semiclassical (plotted upwards, blue online) and quantum-mechanical (plotted downwards, red online) momentum distributions. The horizontal lines in the upper portion of the graph correspond to the heights of the quantum-mechanical peaks.}\label{18comparison}
\end{figure}

\begin{figure}[t]
\includegraphics*[width=\columnwidth]{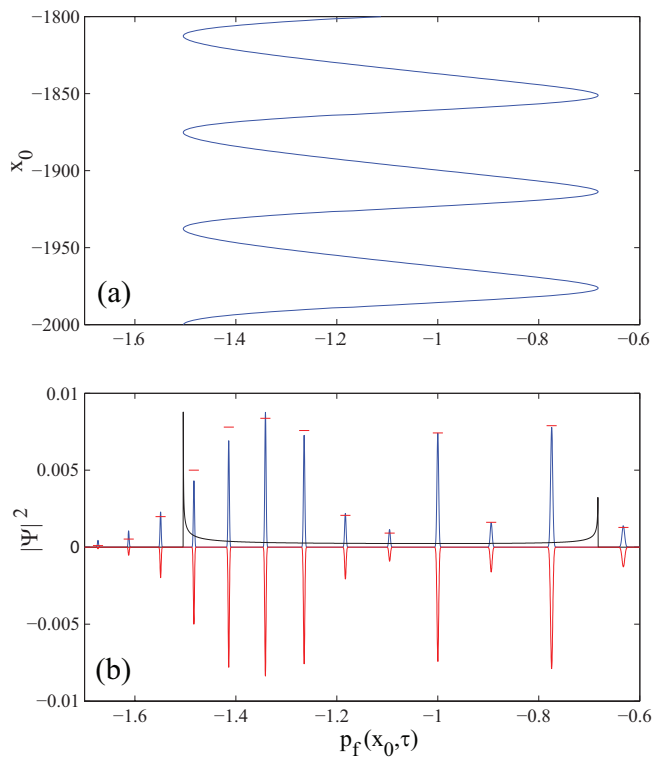}
\caption{(Color Online) Same as Fig.~\ref{18comparison} for $p_0=1.0$.}\label{10comparison}
\end{figure}

In this section, we study three separate scattering cases for identical barrier parameters but different incident momenta: pure transmission, pure reflection, and mixed transmission and reflection. We also compare the full quantum results with the predictions of the semiclassical approach and find relatively good agreement. While there is a large range of possible scattering behaviors that can be studied by adjusting the five input parameters of our model, these three cases capture most of the essential physics.

\emph{Pure Transmission}: The initial Gaussian wave packet is centered at $\bar{x} = -1500$, with $\beta = 300$, with initial momentum $p_{0} = 1.8$, and with barrier parameters $A = 0.5$ and $\omega = 0.1$. This is the same case that was shown earlier in Figs.~\ref{Figure-2}(d),~\ref{fig3regions},~\ref{singles}. This initial momentum corresponds to an energy higher than the maximum amplitude of the barrier. It takes more than fifteen barrier oscillations for the packet to pass over the barrier. There are two branches per cycle, as shown in Fig.~\ref{18comparison}(a). The classically-allowed momentum values range from $p_{f} \approx 1.2506$ to $2.1411$.

A comparison of $\tilde{P}_{SC}^{F}\left(p_{f}\right)$ (plotted upwards, blue online), $\tilde{P}_{Q}^{F}\left(p_{f}\right)$ (plotted downwards, red online) and  $P_{C}^{F}(p_{f})$  (plotted upwards, black online), is shown in Fig.~\ref{18comparison}(b). The semiclassical and quantum-mechanical results can be seen to agree well. The final probability has fifteen peaks within the classical envelope. Both the classical density and interference contribute to the relative heights of peaks. At least two non-negligible classically-forbidden peaks can be seen for momentum values on either side of the classical envelope.  The semiclassical calculation has corrected divergent peaks near momentum turning points by using Airy forms of local wavefunctions (see Appendix~\ref{sec:semiclassical_theory}).

\emph{Pure Reflection:} We employ the same barrier parameters as in the previous case, but use an incident momentum of $p_0=1.0$, which corresponds to an energy equal to the minimum amplitude of the barrier. The barrier undergoes more than twenty-eight oscillations during the time the wave packet is interacting with it. There are two branches per cycle, shown in Fig.~\ref{10comparison}(a), with the classical envelope ranging from $p_{f} \approx -1.5043$ to $-0.6825$.

A comparison of $\tilde{P}_{SC}^{F}\left(p_{f}\right)$ (plotted upwards, blue online), $\tilde{P}_{Q}^{F}\left(p_{f}\right)$ (plotted downwards, red online) and  $P_{C}^{F}(p_{f})$  (plotted upwards, black online), is shown in Fig.~\ref{10comparison}(b), again with good agreement between the semiclassical and quantum-mechanical results. The final-momentum probability has nine peaks within the classical envelope. We see at least three non-negligible peaks for classically-forbidden momentum values less than the minimum of the classical envelope, but only one non-negligible peak for forbidden momentum values greater than the maximum value of the classical envelope. This is because peaks are more closely spaced for large absolute momenta than for small absolute momenta, because they are equally spaced in energy. The exponential decay of the wavefunction again makes the peaks negligible outside the region shown.

\begin{figure}[t]
\includegraphics*[width=\columnwidth]{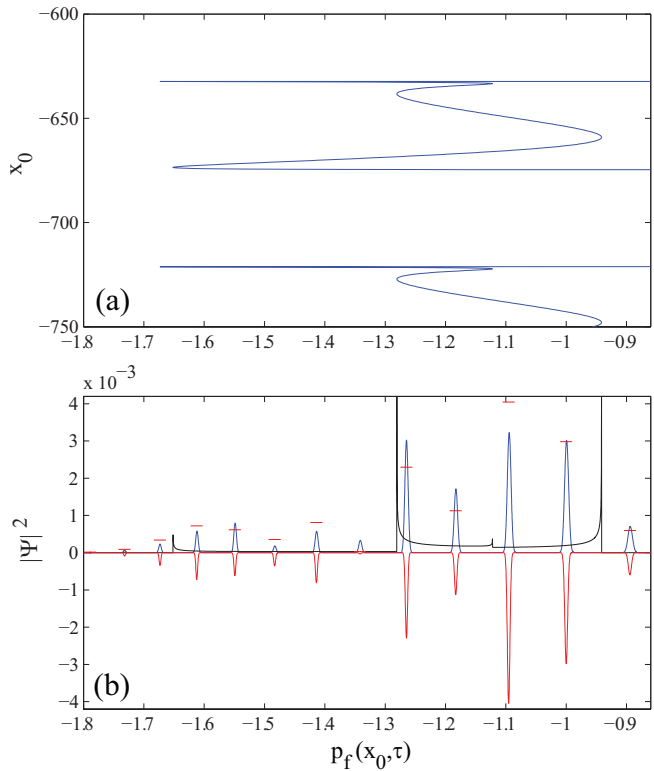}
\caption{(Color Online) (a) Reflected portion of final momentum vs. initial position for the $\omega = 0.1$, $p_{0} = 1.4142$ case. (b) Semiclassical (plotted upwards, blue online), quantum-mechanical (plotted downwards, red online), and classical (plotted upwards, black online) final-momentum probabilities for the reflected portion of the wavepacket.}\label{14comparison_r}
\end{figure}

\begin{figure}[t]
\includegraphics*[width=\columnwidth]{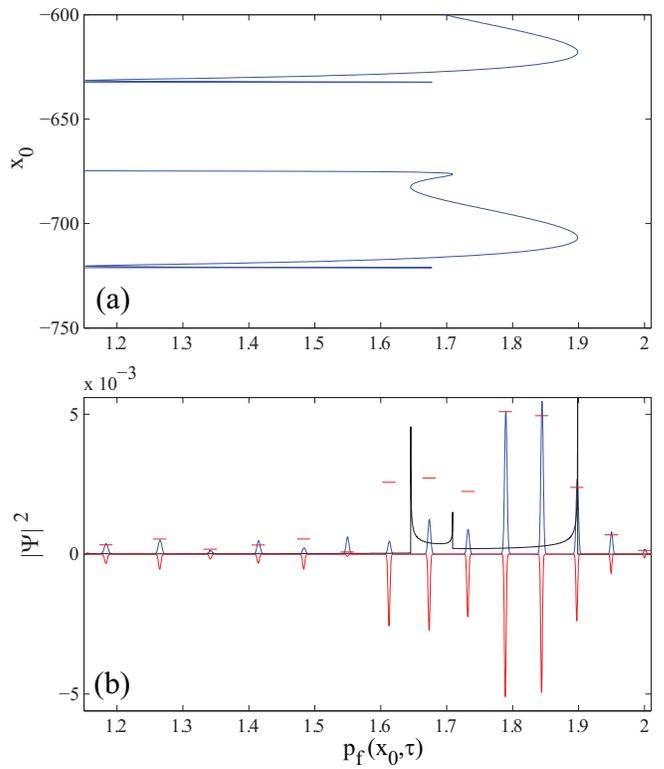}
\caption{(Color Online) (a) Transmitted portion of final momentum vs. initial position for the $\omega = 0.1$, $p_{0} = 1.4142$ case. (b) Semiclassical (plotted upwards, blue online), quantum-mechanical (plotted downwards, red online), and classical (plotted upwards, black online) final-momentum probabilities for the transmitted portion of the wavepacket.}\label{14comparison_t}
\end{figure}

\emph{Mixed Reflection and Transmission:} We implement the same barrier parameters as in the previous cases, but use an incident momentum of $p_0=1.4142$, which corresponds to an energy between the minimum and maximum of the barrier amplitude range. In this case, the wavepacket is partially reflected and partially transmitted. The periodic relationship between final momentum and initial position is more complicated in this case. Figs.~\ref{14comparison_r}(a) and~\ref{14comparison_t}(a) show the reflected and transmitted portions of the trajectory ensemble, respectively. Some classically-allowed final momenta have as many as six interfering trajectories within each cycle. The classical envelope ranges from $p_{f} \approx -1.6730$ to $1.8987$.

Comparisons of the reflected and transmitted portions of  $\tilde{P}^{F}_{SC}\left(p_{f}\right)$ (plotted upwards, blue online), $\tilde{P}_{Q}^{F}\left(p_{f}\right)$ (plotted downwards, red online) and  $P_{C}^{F}(p_{f})$  (plotted upwards, black online)  are shown in Figs.~\ref{14comparison_r}(b) and~\ref{14comparison_t}(b), respectively. Every extremum in the $p_{f}\left(x_{0},t_{f}\right)$ graph gives a ``turning point'' or caustic, at which $P_{C}\left(p_{f}\right)$ diverges. The classical amplitude is markedly higher for larger momentum values in both the reflected and transmitted portions of the wave packet; consequently, the semiclassical and quantum-mechanical final-momentum distributions have their largest peaks in these regions. Agreement between semiclassical and quantum methods is less precise in this case, particularly where turning points are close together.

\section{Proposed Experiment}\label{sec:experiment_section}
The theoretical predictions of the previous sections can be tested experimentally with the macroscopic wavefunction of a BEC serving as the atomic wave packet. While the BEC does not have to be strictly 1D, the use of a highly elongated BEC, confined in an optical dipole trap, simplifies the experiment. Furthermore, the BEC should be non-interacting since collisions between particles are not included in our calculations. A non-interacting BEC can be produced by employing a magnetic Feshbach resonance.
A number of alkali atoms, such as $^{85}$Rb \cite{WiemanCornell_Rb85Feshbach_PRL1998} and $^{39}$K \cite{Inguscio_PRL2007}, have
been cooled to quantum degeneracy and also have a Feshbach ``zero'', a magnetic field which produces a null
scattering length due to a nearby Feshbach resonance. In the non-interacting limit the Thomas-Fermi approximation no longer applies, and a harmonically confined BEC has the Gaussian wavefunction of the trap ground state.

\begin{figure}
\includegraphics[height=0.27\textheight]{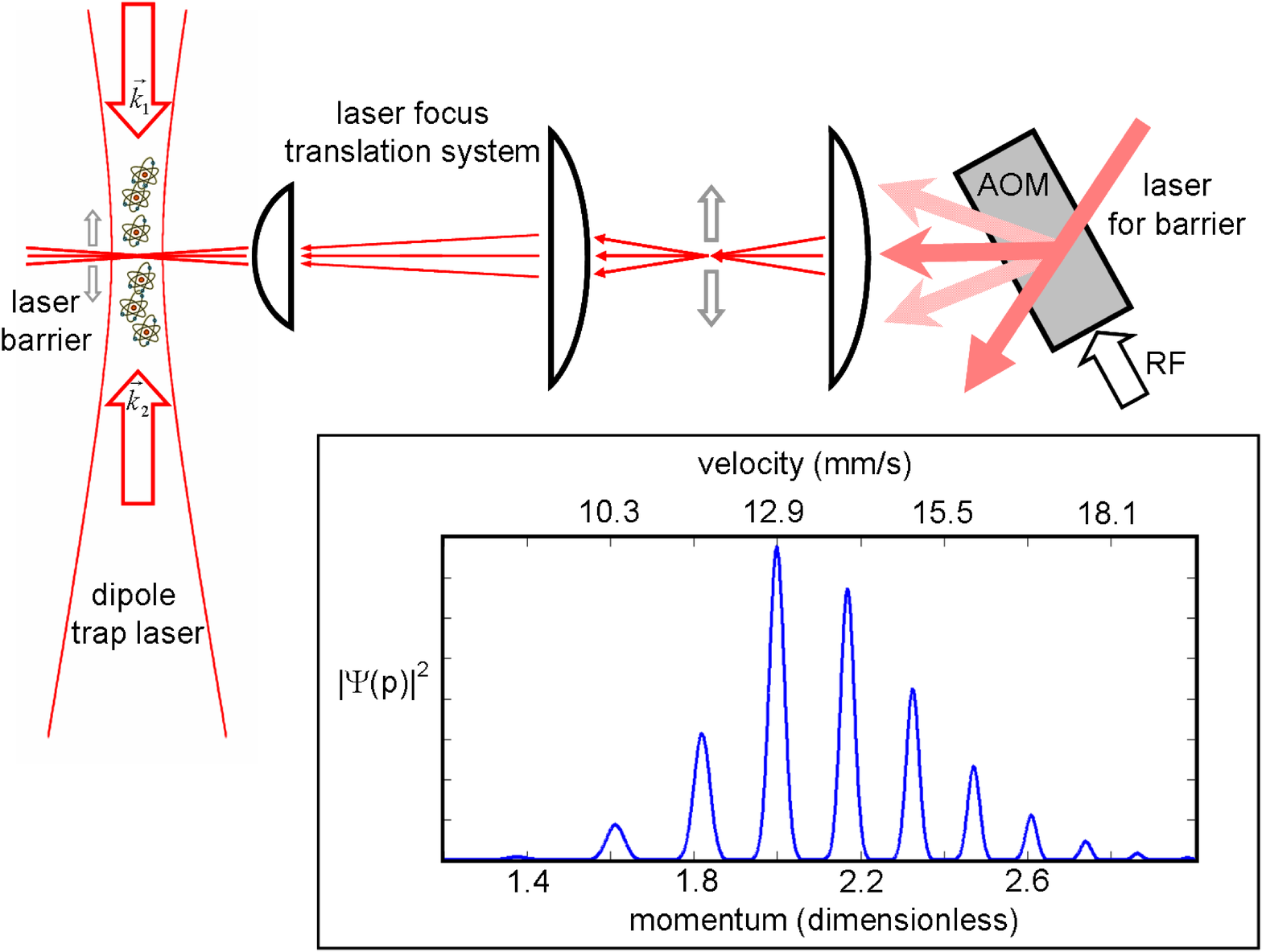}
\caption{\label{ExperimentalImplementation_figure}(Color Online) Proposed experimental implementation of oscillating barrier scattering. An AOM generates a barrier laser beam whose deflection angle is controlled by the RF drive frequency. The beam rotation is converted to a translation (grey arrows) by a lens, which also focuses the beam to produce a narrow dipole barrier. A combination of lenses then inserts the laser barrier onto the BEC (atom symbols). $\vec{k}_1$ and $\vec{k}_2$, and associated arrows, indicate the wavevectors for the Bragg-Raman spectroscopy laser beams. The inset shows a sample momentum and velocity distribution for the proposed Bragg-Raman spectroscopy experiment with a $^{39}$K BEC.}
\end{figure}

\begin{table}[b]
\caption{\label{table:experiment_parameters} Table of proposed experiment parameters.}
\begin{ruledtabular}
\begin{tabular}{lc}
\textrm{Parameter}&
\textrm{Value}\\
\colrule
atomic state & $|F=1,m_F=+1\rangle$ state of $^{39}$K\\
Feshbach zero & 350 G\\
BEC width $\beta$ & 10 $\mu$m\\
BEC velocity & 12.9 mm/s\\
barrier width $\sigma$ & 2.5 $\mu$m\\
barrier amplitude $U_0$ & 197 nK\\
barrier mod. ampl. A & 1\\
barrier mod. frequency $\omega$ & $2\pi\times1.4 kHz$\\
\end{tabular}
\end{ruledtabular}
\end{table}

The elongated BEC provides a wide, quasi-1D, Gaussian wavepacket, while a tightly focused blue-detuned laser serves as an optical dipole barrier with Gaussian shape, and with amplitude proportional to the laser intensity. Instead of launching the atoms towards the barrier,
the opposite is more convenient in an experiment: The dipole barrier is swept through the stationary BEC, so that in the reference frame of the moving
barrier the theoretical treatment still applies. Figure \ref{ExperimentalImplementation_figure} shows an optical circuit for
generating a translating laser barrier using an acousto-optic modulator (AOM): the amplitude and frequency of the radio-frequency (RF)
drive for the AOM control the amplitude and position, respectively, of the barrier.

The momentum components of the reflected and transmitted BEC wavefunction will generally be too close together to be distinguishable by time-of-flight imaging. Instead, Bragg spectroscopy \cite{Ketterle_Bragg_PRL1999} can be used to measure the
velocity distribution of the scattered atomic packet. Bragg spectroscopy is performed by briefly shining two laser beams on the scattered atoms, as shown in Fig.~\ref{ExperimentalImplementation_figure}. When the two lasers are detuned from each other by a frequency $\delta=\delta_0+(\vec{k}_1-\vec{k}_2)\cdot\vec{v}$, where $\vec{k}_1$ and $\vec{k}_2$ are the wavevectors of the two incident Bragg beams, then only atoms with velocity $\vec{v}$ are given a two-photon momentum kick $\hbar(\vec{k}_1-\vec{k}_2)$. The energy imparted to an atom at rest by two-photon recoils determines the base detuning $\delta_0=\hbar k^2/(2m)$.  The kicked atoms are spectroscopically tagged with a Raman process \cite{KasevichChu_Raman_PRL1991} that changes their hyperfine level, accomplished by adding the hyperfine ground level splitting to the base detuning $\delta_0$. The Raman-selected atoms are detected by absorption or fluorescence imaging on the D2 line cycling transition.

We summarize the main parameters of the proposed experimental implementation in table \ref{table:experiment_parameters}. We consider a BEC of $^{39}$K atoms in the $|F=1, m_F=+1\rangle$ hyperfine ground state, which has a vanishing s-wave scattering length at 350 G \cite{Inguscio_PRL2007}. A red-detuned optical dipole trap produced by a 1 W 1064 nm laser focused to a 1/e$^2$ diameter
of about 120 $\mu$m will confine the BEC with a Gaussian density profile and an axial width of $\beta=40\equiv10 \mu$m.  A blue-detuned Gaussian
barrier can be produced with a 532 nm laser focused to a radius of $\sigma=10\equiv2.5 \mu$m (waist radius of 5 $\mu$m) with a barrier amplitude of $U_0=1\equiv197$ nK. Translating this barrier at a velocity of 12.9 mm/s (corresponding to an incident momentum of $p_0=2$ for particles of mass $m=1=6.5\times 10^{-26}$ kg), while modulating it at $\omega_{barrier}=0.35\equiv2\pi\times$1.4 kHz with a modulation strength of $A$=1, produces a purely transmitted wavepacket with the final momentum distribution shown in the inset of Fig.~\ref{ExperimentalImplementation_figure}. The velocity peaks of the distribution have a half width at half maximum of $\Delta v\approx$0.1 mm/s, determined by the axial extent of the BEC. This velocity spread requires a laser frequency difference stability on the order of $2\Delta v/\lambda\approx$250 Hz ($\lambda=767$ nm for $^{39}$K), which is within the practical resolution of Bragg spectrocopy \cite{ThywissenAspect_Bragg_PRL2003}. Furthermore, we note that the axial confinement of the BEC does not play a significant role, since the trap has an axial oscillation frequency of $f_{axial}\approx$1 Hz, which is considerably slower than the time scale of the scattering process.

\section{Conclusion}\label{sec:conclusion_section}
In summary, we have studied scattering from an amplitude-modulated Gaussian barrier, and determined the final momentum-space probability distributions using classical, semiclassical, and quantum formalisms. We find that classical mechanics defines the boundaries of a classically-allowed region of final momenta. Quantum calculations show: (i) the probability that particles end with momentum outside the classically-allowed region is small; (ii) the momentum distribution is peaked at momenta consistent with Floquet's theorem; (iii) the heights of the Floquet peaks vary widely and seemingly erratically. Semiclassical calculations show that (a) for any final momentum inside the classically-allowed region, many classical paths arrive; (b) interference of waves propagating along these paths produces peaks consistent with Floquet theory, and determines their heights. Specifically, inter-cycle interference leads to discrete final momentum states, while intra-cycle interference determines the peak heights. Finally, momentum-space tunneling leads to diffractive population of momenta beyond the classically-allowed bounds.

The semiclassical and full quantum propagation formalisms employed in this work are well suited for studying scattering from a turnstile pumping potential formed from two separated barriers, amplitude-modulated $\pi/2$ out of phase from each other. Classically, such a potential displays strong signatures of chaos, with quantum dynamics well suited to the type of semiclassical treatment developed in this paper. Such a treatment is essential for understanding the quantum and classical aspects of particle pumping in a turnstile pump, since interference and tunneling can be selectively included. Moreover, the scattering theories developed in this work can also be extended to examine spatial tunneling through narrower barriers, and scattering from a potential well.

\section{Acknowledgements}
K.K.D. and A.J.P. acknowledge support of the NSF under Grant No. PHY-0970012. K.A.M. acknowledges NSF support via Grant No. PHY-0748828. JBD and TAB acknowledge support from NSF Grant No. PHY-1068344.

\appendix

\section{semiclassical Analysis}\label{sec:semiclassical_theory}

We give here details and derivation of the semiclassical formulas used in Sec.~\ref{sec:semiclassical_description}. Most of the theory is similar to methods we have used in earlier papers \cite{MasFed,delos863,schwaldelos,spellabunch,haggplus3,haggdelos,cdsanddelos1,cdsanddelos2,dudelos,dudelos2,dudelos3}, but some aspects of the present system are different. In most of our earlier work, we have studied stationary fixed-energy systems; only \cite{haggplus3,haggdelos,cdsanddelos1,cdsanddelos2} dealt with time-dependent potentials. In the present case, the initial and final conditions are, from semiclassical perspectives, a little unusual. At the final time, we want a semiclassical approximation in momentum space. However, at the initial time, we cannot use a semiclassical approximation in momentum space, though we can in configuration space. Furthermore, the sum over cycles of the oscillating barrier is different from previous work.

\subsection{Local Wavefunction}\label{subsubsec:semiclassical local}
Recall that we have an oscillating Gaussian barrier with a wave packet approaching from the left. At an initial time $t_{0}$, the wavefunction for $x\ll 0$ (far to the left of the barrier), is given by
\begin{align}
\Psi_{0}(x,t_{0})=F(x)e^{i\left(p_{0}x-E_{0}t_{0}\right)/\hbar},
\end{align}
where $F(x)$ is a function describing the envelope of the initial packet in $(x,t)$ space. We include time and energy as canonical variables, expanding the phase space for the system. For reasons that will become clear, one regards $t$ as a canonical momentum, and $E$ as a canonical coordinate, $q=(x,E)$ and $p=(p,t)$.

\begin{figure}[t]
\includegraphics*[width=\columnwidth]{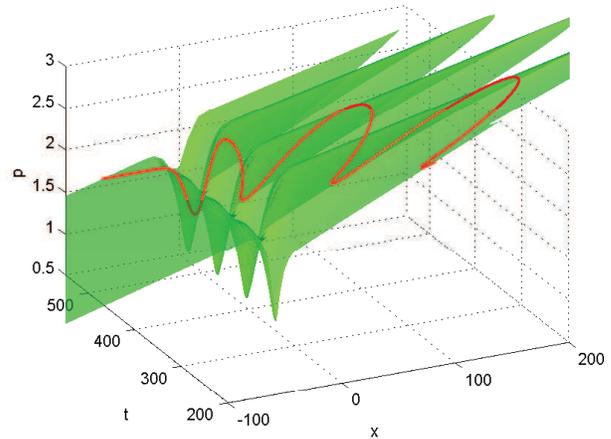}
\caption{(Color Online) Typical Lagrangian manifold for this system. The solid line (red online) shows a slice at a constant time.}\label{Single-Barrier-LM}
\end{figure}

Then defining an effective Hamiltonian, $\mathscr{H}$, given by
\begin{align}
\mathscr{H}=\frac{p^2}{2m}+U(x,t)-E,
\end{align}
the equations of motion are
\begin{subequations}
\begin{alignat}{5}
\frac{dx}{d\tau} &=&\frac{\partial \mathscr{H}}{\partial p} &=&\frac{\partial H}{\partial p}\\
\frac{dp}{d\tau} &=&-\frac{\partial \mathscr{H}}{\partial x} &=&-\frac{\partial H}{\partial x}\\
\frac{dE}{d\tau} &=&\frac{\partial\mathscr{H}}{\partial t}&=&\frac{\partial U}{\partial t}\label{dedtau}\\
\frac{dt}{d\tau} &=&-\frac{\partial \mathscr{H}}{\partial E}&=&1\label{dtdtau}\\
\frac{dS}{d\tau}&=&p\frac{dx}{d\tau}&-&E\frac{dt}{d\tau}\\
\frac{d \tilde{S}}{d\tau}&=&-x\frac{dp}{d\tau}&-& E\frac{dt}{d\tau}\label{hah2}
\end{alignat}
\end{subequations}
where $\tau$ is a ``timelike'' progress variable along the trajectories, and is related to $t$ in the Schr\"{o}dinger Equation via $\tau=t_{0}+t$. We call $S$ the classical action along the trajectory, and $\tilde{S}$ can be thought of as a ``momentum-space action'' along the trajectory. The form of equations \eqref{dedtau} and \eqref{dtdtau} justify the indentification of $E$ as a canonical coordinate and $t$ as its conjugate momentum.

We want to compute the probability that the particles end with a given final momentum, using the momentum-space wavefunction $\tilde{\Psi}\left(p,t\right)$. Therefore, we want a semiclassical approximation in momentum space. However, since we have chosen an initial distribution with very small momentum spread, the initial wavefunction in momentum space is nearly a delta function, which cannot be described by a semiclassical approximation. Therefore, in order to calculate the desired momentum-space wavefunction, we start our calculation in $(x, t)$ space, and later transform to $(p, t)$ space.

The first step in constructing a semiclassical wavefunction is is to propagate trajectories from a line of initial conditions. We choose the line of initial conditions to have a constant starting time $t_0=0$, variable starting position $x$ covering the domain of the initial packet, and a fixed initial momentum $p_0$. The resulting trajectories sweep out a two-dimensional surface called a Lagrangian manifold in the four-dimensional $\left(x,p,E,t\right)$ phase space. A typical Lagrangian manifold for this system is shown in Fig.~\ref{Single-Barrier-LM}.

Integration of trajectories with respect to $\tau$ gives a relationship between $(x_{0}, \tau)$ and $(z, t)$, where $z$ is any dynamical variable $x, p, E$, $S$, or $\tilde{S}$. From our choice of $t_{0}=0$, $t$ is simply equal to $\tau$, and $x$ is the point at which the trajectory arrives at time $t=\tau$. We may think of each of these quantities as a function of the initial variable $x_{0}$ and the progress variable $\tau$, e.g., $x(x_{0}, \tau), p(x_{0}, \tau)$, etc.

\begin{figure}[t]
\includegraphics*[width=\columnwidth]{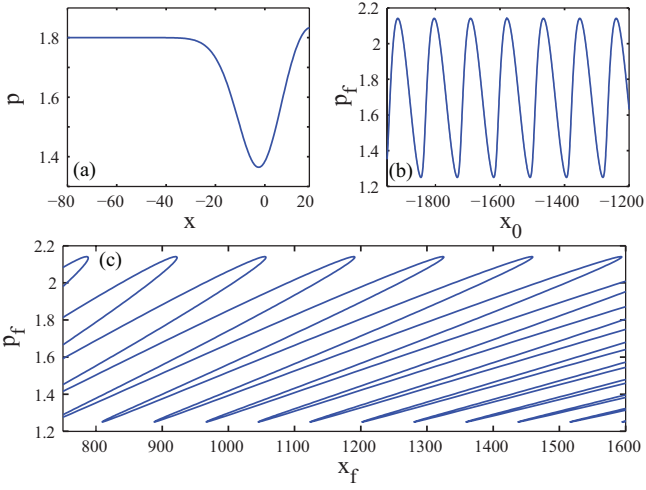}
\caption{(a) Slice of Lagrangian manifold at small time. (b) Periodic final momentum as a function of initial position. (c) Final momentum,  $p_{f} = p\left(x_{0},\tau_{f}\right)$, as a function of final position, $x_{f} = x\left(x_{0},\tau_{f}\right)$. This corresponds to the final-time slice of the Lagrangian manifold.}\label{Single-Barrier-Figtommy1}
\end{figure}

We define a Jacobian,
\begin{align}
J(x_{0}, \tau)&=\det\left(\frac{\partial (x,t)}{\partial (x_{0},\tau)}\right)=\frac{\partial x}{\partial x_{0}}
\end{align}
with $J_{0}=J\left(x_{0},0\right)=1$. This Jacobian is a single-valued function of $(x_{0}, \tau)$. For $\tau$ not too large (and $x$ not too far from $x_{0}$) there is an invertible relationship between $(x_{0}, \tau)$ and $(x, t)$; i.e., we may consider $(x_{0}, \tau)$ as a function of $(x, t)$. With this relationship, we may also consider the position-space action $S$ and Jacobian $J$ to be functions of $(x, t)$,

\begin{figure}[t]
\includegraphics*[width=\columnwidth]{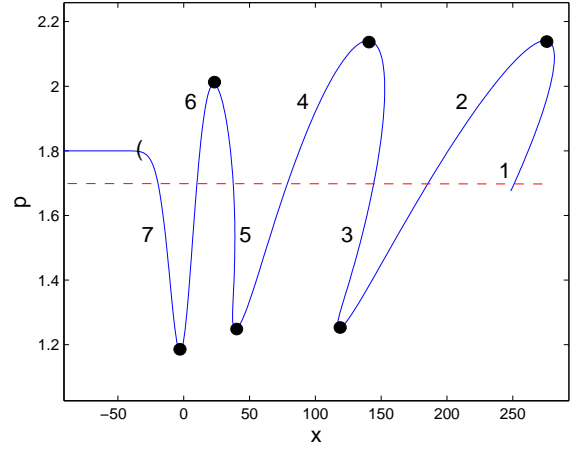}
\caption{(Color Online) Slice of Lagrangian manifold at an intermediate time. The numbers correspond to intermediate-time slices of different momentum charts, which are separated by local extrema in the function $p =  \mathfrc{P}_{\mbox{   }}(x, t)$ for fixed time, denoted by large circles. For every given momentum (e.g., the dashed line), there are many corresponding values of $x$.}\label{Single-Barrier-mediumFigtommy1}
\end{figure}

\begin{alignat}{5}
S(x_{0}, \tau)&=&S(x_{0}(x, t), \tau(x, t))&=&\mathcal{S}(x, t)\n\\
J(x_{0}, \tau)&=&J(x_{0}(x, t), \tau(x, t))&=&\mathcal{J}(x, t)
\end{alignat}
We may use these functions in the primitive semiclassical approximation for the $(x, t)$ space wavefunction
\begin{align}
\Psi_{SC}(x, t)=\Psi_{0}(x_{0}, \tau=0)\left|\frac{\mathcal{J}_{0}}{\mathcal{J}(x, t)}\right|^{1/2}e^{i \mathcal{S}(x, t)/\hbar}\label{psiconfig},
\end{align}
where $(x_{0},\tau)$ are considered to be functions of $(x,t)$. The initial Maslov index has been set equal to zero, and
\begin{align}
&\Psi_{0}(x_{0}, \tau=0)=F(x_{0})e^{ip_{0}x_{0}/\hbar},
\end{align}
where $(x_{0}, \tau)$ are again considered as functions of $(x,t)$.

As the trajectories are propagated forward in $\tau$, they come to the barrier region, where $p$ is no longer constant, and we may use $(p, t)$ locally as independent variables to describe the Lagrangian manifold, as shown in Figs.~\ref{Single-Barrier-LM},~\ref{Single-Barrier-Figtommy1}(a) and~\ref{Single-Barrier-Figtommy1}(c), and Fig.~\ref{Single-Barrier-mediumFigtommy1}.

A ``momentum chart'' is a region of the Lagrangian manifold that has a diffeomorphic projection to momentum space, $(p, t)$. In Fig.~\ref{Single-Barrier-mediumFigtommy1}, a constant-time slice of the Lagrangian manifold is shown. For each value of $p$, there are many corresponding values of $x$; each can be regarded as a ``branch'' of a multivalued function, and each is a constant-time slice of a momentum chart.

We transform to the momentum-space wavefunction via
\begin{align}
\tilde{\Psi}(p,t)=&(2 \pi i \hbar)^{-1/2}\int{\Psi_{\scriptscriptstyle{sc}}(x,t) e^{-i px/\hbar}dx}.\label{jshdbjhb}
\end{align}
We evaluate the integral for the part of the wavefunction that corresponds to the initial momentum chart by using the stationary phase approximation. We use the function $ \mathfrc{P}_{\mbox{   }}(x, t)=\partial S/\partial x$ to describe the Lagrangian manifold, and $p$ is the independent variable in $\tilde{\Psi}(p, t)$. When we substitute the semiclassical approximation \eqref{psiconfig} into \eqref{jshdbjhb}, each classically-allowed $p$ has a stationary phase point, $\hat{x}$, where $p= \mathfrc{P}_{\mbox{   }}(\hat{x}, t)$, i.e., where the line $p=$ constant intersects the Lagrangian manifold, as shown in Fig.~\ref{Single-Barrier-mediumFigtommy1} for $p = 1.7$. In evaluating the integral, we also make use  of the momentum-space action, defined in Eq. \eqref{hah2}, and define a momentum-space Jacobian

\begin{align}
\tilde{J}(x_{0}, \tau)=&\det\left(\frac{\partial(p,t)}{\partial(x_{0}, \tau)}\right) = \frac{\partial p}{\partial x_{0}}.
\end{align}
The locally invertible relationship between $(p, t)$ and $(x_{0}, \tau)$ allows us to consider $\tilde{S}(x_{0}, \tau)$ and $\tilde{J}(x_{0}, \tau)$ to be functions of $(p, t)$, i.e.,
\begin{align}
\tilde{S}(x_{0}(p, t), \tau(p, t))&=\tilde{\mathcal{S}}(p, t)\n\\
\tilde{J}(x_{0}(p, t), \tau(p, t))&=\tilde{\mathcal{J}}(p, t).
\end{align}
With these definitions, the stationary phase approximation in the initial momentum chart yields
\begin{align}
\tilde{\Psi}_{1}(p,t)&=F(x_{0}^{1}(p, t))e^{i\left(\tilde{\mathcal{S}}_{1}(p, t)/\hbar-\pi/2\right)}\left|\tilde{\mathcal{J}}_{1}(p, t)\right|^{-1/2}\label{pfn}
\end{align}

Generally, for every momentum chart of the Lagrangian manifold, there is a comparable term contributing to the momentum-space wavefunction. We write the local, primitive form of the momentum-space wavefunction for each momentum chart as
\begin{align}
\begin{split}
\tilde{\Psi}_{j}(p,t)&=F(x_{0}^{j}(p, t)) \left|\frac{1}{\tilde{\mathcal{J}}_{j}(p, t)}\right|^{1/2}\\
&\times\exp{\left(\frac{i \tilde{\mathcal{S}}_{j}(p, t)}{\hbar}-\frac{i \pi \tilde{\mu}_{j}}{2}\right)},
\end{split}\label{gen}
\end{align}	
where $\tilde{\mu}_{j}$ is the Maslov index for the given momentum chart.

\subsubsection{Maslov Index}\label{subsubsec:semiclassical maslov}
Here, we state the rule for the Maslov index for each momentum chart. As indicated in Fig.~\ref{Single-Barrier-mediumFigtommy1}, momentum charts are separated by momentum turning points, which are extrema of locally-defined functions $p =  \mathfrc{P}_{\mbox{   }}(x, t)$ for fixed $t$, i.e., points where $\partial  \mathfrc{P}_{\mbox{   }}(x, t)/ \partial x = 0$.

Each time any path on the Lagrangian manifold passes through a momentum turning point, the Maslov index changes. In Fig.~\ref{Single-Barrier-Figtommy1}(c), we show a slice of the Lagrangian manifold at the final time $t_f$. If we take any two points on this slice of the manifold, they can be connected by a path on this slice. At each point that the path passes through a momentum turning point, the Maslov index changes by $\pm1$, and we use the following rule to determine the increment. This rule applies if the $\left(x,p\right)$ plane is drawn in the most usual way, with $x$ increasing to the right and $p$ increasing upward.
When the path passes through a momentum turning point that separates the $i^{th}$ momentum chart from the $j^{th}$ momentum chart, then
\begin{subequations}
\begin{align}
\tilde{\mu}_{j}&=\tilde{\mu}_{i}+1, &\mbox{if the path curves right (CW)}\label{rightmu}\\
\tilde{\mu}_{j}&=\tilde{\mu}_{i}-1, &\mbox{if the path curves left (CCW)}\label{leftmu}
\end{align}
\end{subequations}
where CW and CCW denote clockwise and counter-clockwise, respectively.

The $\exp\left(-i\pi/2\right)$ term in the primitive wavefunction for the first momentum chart, \eqref{pfn}, corresponds to $\tilde{\mu}_{j}=1$ in \eqref{gen}. All other Maslov indices for the remaining momentum charts are constructed relative to it, using \eqref{rightmu} and \eqref{leftmu}.

For the two paths shown in Figs.~\ref{Single-Barrier-Figtommy1}(c) and~\ref{Single-Barrier-mediumFigtommy1}, moving from left to right, the Maslov index increases at every maximum, and decreases at every minimum.

\subsubsection{Corrections Near Momentum Turning Points}\label{subsubsec:semiclassical correction}
The primitive semiclassical wavefunction diverges at momentum turning points, where where $\tilde{\mathcal{J}}_{b}(p, t)$ vanishes. To correct this, we construct an alternative way of writing the primitive wavefunction, which will be valid near momentum turning points in classically-allowed regions. We then match this form of the wavefunction to the Airy function and its derivative, in order to extend the semiclassical approximation into classically-forbidden regions \cite{child}.

We start by adding the primitive forms of the wavefunction, \eqref{gen}, for two successive momentum charts, and we denote this wavefunction $\tilde{\Psi}_{m+n}(p,t)$. We introduce the following notation
\begin{subequations}
\begin{align}
A(p, t)&=\left|\tilde{\mathcal{J}}(p, t)\right|^{-1/2}\\
\Delta \tilde{\mathcal{S}}(p,t)  &= \tilde{\mathcal{S}}_{n}(p,t) - \tilde{\mathcal{S}}_{m}(p,t)\label{firstbars}\\
\tilde{\mathbb{S}}(p,t) &=\left[\tilde{\mathcal{S}}_{n}(p,t) + \tilde{\mathcal{S}}_{m}(p,t)\right]/2\label{sbar}\\
\Delta A(p,t)&=A_{n}(p,t) - A_{m}(p,t) \label{deltaa}\\
\mathbb{A}(p,t) &=\left[A_{n}(p,t)+A_{m}(p,t)\right]/2\label{abar}\\
\Delta F(x_{0}(p, t)) &=F_{n}(x_{0}(p, t)) - F_{m}(x_{0}(p, t))\label{deltaf}\\
\mathbb{F}(x_{0}(p, t)) &=\left[F_{n}(x_{0}(p, t)) + F_{m}(x_{0}(p, t))\right]/2,\label{lastbars}
\end{align}
\end{subequations}
where the $m$ and $n$ subscripts denote the momentum chart with the lower and higher Maslov index, respectively. We use these definitions to write
\begin{align}
\begin{split}
&\tilde{\Psi}_{m+n}(p,t)=2\exp\left({\frac{i\tilde{\mathbb{S}}(p,t)}{\hbar}} - \frac{i \tilde{\mu}_{m}\pi}{2}\right)\\
&\times \Bigg\{\left(\mathbb{A}\mathbb{F} + \frac{\Delta A \Delta F}{4}\right)
\mbox{e}^{-i\pi/4}\sin{\left(\frac{\Delta\tilde{\mathcal{S}}(p,t)}{2\hbar}+\frac{\pi}{4}\right)}\Bigg.\\
&   + \Bigg.\left(\frac{\mathbb{A}\Delta F}{2}+\frac{\Delta A \mathbb{F}}{2}\right)
\mbox{e}^{-i3\pi/4}\cos{\left(\frac{\Delta\tilde{\mathcal{S}}(p,t)}{2\hbar}+\frac{\pi}{4}\right)}\Bigg\}
\end{split}\label{sumofterms}
\end{align}

We match the separate terms of \eqref{sumofterms} to the first-order asymptotic forms of the Airy function and its derivative, respectively, so that we may write \eqref{sumofterms} as
\begin{align}
\begin{split}
\tilde{\Psi}_{m+n}(p,t)&= C\left(p,t\right) \mbox{Ai}(-z(p,t)) \\
&+ D\left(p,t\right) \mbox{Ai}'(-z(p,t)),\label{impairys}
\end{split}
\end{align}
where
\begin{subequations}
\begin{align}
&C= \frac{2\exp\left(i\left(\frac{\tilde{\mathbb{S}}}{\hbar} - \frac{\tilde{\mu}_{m}\pi}{2}-\frac{\pi}{4}\right)\right) \left[\mathbb{A}\mathbb{F} + \frac{\Delta A \Delta F}{4}\right]}{\pi^{-1/2} \left(z\left(p, t\right)\right)^{-1/4}}\label{cone}\\
&D= \frac{-2\exp\left(i\left({\frac{\tilde{\mathbb{S}}}{\hbar}} - \frac{ \tilde{\mu}_{m}\pi}{2}-\frac{3\pi}{4}\right)\right)
\left[\frac{\mathbb{A}\Delta F}{2}+\frac{\Delta A \mathbb{F}}{2}\right]}{\pi^{-1/2} \left(z\left(p, t\right)\right)^{1/4}}\label{ctwo}\\
&z(p,t) = \left(\frac{3 \Delta\tilde{\mathcal{S}}(p,t)}{4\hbar}\right)^{2/3}\label{zpx}
\end{align}
\end{subequations}
We use wavefunctions of the form of \eqref{impairys} in the classically-allowed regions near momentum turning points, where \eqref{gen} is not valid.

\subsubsection{Classically-Forbidden Regions}\label{subsubsec:semiclassical forbidden}
One can show that if the momentum turning points are quadratic maxima or minima, the following functions vary linearly with $p$ near the turning point $\hat{p}$
\begin{subequations}
\begin{align}
[\Delta \tilde{\mathcal{S}}(p,t)]^{2/3} &\propto \left(p - \hat{p}\right) \label{hereone}\\
\tilde{\mathbb{S}}(p,t) + \tilde{\mathbb{S}}(\hat{p},t) &\propto \left(p - \hat{p}\right)\\
\left[\mathbb{A}(p,t)\right]^{-4} &\propto \left(p - \hat{p}\right)\\
\left[\Delta A(p,t)\right]^{4} &\propto \left(p - \hat{p}\right)  .\label{heretwo}
\end{align}
\end{subequations}
We continue these quantities into the classically-forbidden regions using these linear approximations. To obtain values for $F(x_{0}(p, t))$ in these regions, we extrapolate $x_{0}$ into the classically-forbidden regions, and use it to evaluate $F(x_{0}(p, t))$. This extrapolation yields complex values of $x_{0}$.

\subsection{Global Wavefunction}\label{subsubsec:semiclassical global}

We denote as ``branches'' the regions separated by momentum turning points in $p(x_{0}, \tau_{f})$, i.e., regions separated by points where $\partial p(x_{0}, \tau_{f})/\partial x_{0} = 0.$  We define a ``cycle'' as one barrier oscillation, i.e., one period of $p(x_{0}, \tau_{f})$.

We want to construct a final wavefunction that is valid in both classically-allowed and classically-forbidden regions. We have seen that each momentum chart contributes a term to the final wavefunction, so our first step is to construct all local wavefunctions.

We will illustrate the steps necessary to construct the final wavefunction for the simplest case, like that shown in Fig.~\ref{fig3regions}(a), which contains two branches per cycle. We must determine the regions of validity of the two forms of the wavefunction, Eqs. \eqref{gen} and \eqref{impairys}, for all branches. Due to the periodicity of final momentum and initial position, we can do this for a single cycle only, as Eqs. \eqref{gen} and \eqref{impairys} are valid in the same regions for the $i^{th}$ branch within every cycle. Further consequences of this periodicity are discussed in App.~\ref{sec:kevin}.

We choose the cycle spanning branches $\left(a,b,c\right)$ in Fig.~\ref{fig3regions}(a). We start with branches $a$ and $b$, and construct the primitive form of the wavefunction by adding Eq. \eqref{gen} for the two branches. We then construct $\tilde{\Psi}_{a+b}(p,t)$ via Eq. \eqref{impairys}. These two forms of the wavefunction are valid in different but overlapping regions, and we compare the two to determine the region of validity for each. This comparison shows that the Airy form is valid in regions $D$ and $E$ in Fig.~\ref{fig3regions}(a) ($p$ $\scriptstyle{\lesssim}$ $1.66$). In region $D$ ($1.36$ $\scriptstyle{\lesssim}$ $p$ $\scriptstyle{\lesssim}$ $1.66$), where both forms of the wavefunction are valid, we use a switching function that varies between 0 and 1 to weight each form, and use a linear combination of the two. We then use
\begin{align}
\begin{split}
\tilde{\Psi}_{a+b}(p,t) =&f_{1}\left(p\right)\left[\mbox{Airy form}\right] \\
&+ \left(1-f_{1}\left(p\right)\right)\left[\mbox{Prim. form}\right],\label{weight2}
\end{split}
\end{align}
as the local wavefunction for branches a and b in regions $C$, $D$, and $E$, where $f_{1}$ is the switching function; $f_{1}\rightarrow 0$ at the boundary between regions $C$ and $D$. ``Airy form'' and ``Prim. form'' in Eq. \eqref{weight2} refer to $\tilde{\Psi}_{a+b}(p,t)$ calculated via Eqs. \eqref{impairys} and \eqref{gen}, respectively.

We repeat this process for branches $b$ and $c$, and find that the Airy form of this wavefunction, $\tilde{\Psi}_{b+c}(p,t)$, is valid in regions $A$ and $B$ ($p$ $\scriptstyle{\gtrsim}$ $1.77$). Both forms of the wavefunction are valid in region $B$ ($1.77$ $\scriptstyle{\lesssim}$ $p$ $\scriptstyle{\lesssim}$ $2.01$).We use a switching function in region $B$ to weight each form of the wavefunction, and use a linear combination of the two. We use primitive semiclassical wavefunctions for all branches in region $C$ ($1.66$ $\scriptstyle{\lesssim}$ $p$ $\scriptstyle{\lesssim}$ $1.77$).

With knowledge of where each branch's primitive and Airy forms of the local wavefunction may be used, one may construct a final wavefunction, which is a linear combination of all local wavefunctions. For cases with more than two branches per cycle, a more elaborate version of the same process is used.

\section{Semiclassical Implications of Periodicity}\label{sec:kevin}

An initial wavefunction that is long in position space needs many
oscillation cycles to pass through the barrier region.
Semiclassically, this means that the summation of the primitive
wavefunction $\tilde{\Psi}_j(p, t)$ (Eq.\eqref{gen}) over the
momentum charts $j$ involves a sum over trajectories with initial
$x_0$ values extending over numerous oscillation cycles of $p(x_0,
\tau_f)$, as seen in Fig.~\ref{Single-Barrier-Figtommy1}(b).  This
creates interference of trajectories belonging to different cycles of oscillation. This inter-cycle interference constructively enhances
final momentum values satisfying
$\Delta E = \hbar \omega$, consistent with Floquet theory.  Here we
clarify how this constraint arises semiclassically and derive
explicit formulas for the resulting momentum-space wavefunctions.
The following discussion refers to the classically-allowed regions, but its validity could be
extended to include regions near turning points, and classically-forbidden regions, by using the appropriate Airy forms of local wavefunctions.

Let $L$ denote the initial interval of $x$-values over which the
initial wave packet is defined.  We consider all
those trajectories ending with a given value of $p_{f}$ and beginning
with any initial $x_0$ in $L$.  We further restrict attention to
trajectories whose final point $x\left(x_{0},\tau_{f}\right)$ is sufficiently far outside the
barrier region that the potential is essentially flat.  This is
appropriate when most of the wave packet has either reflected from or
passed through the barrier region.  Now, we choose some interval $I$
of length $p_0 T/m$, corresponding to one oscillation period $T = 2\pi/\omega$, within $L$.  Label all those trajectories ending at $p_f$ which have $x_0$
inside $I$ with an index $b$ as above; i.e. the initial position for
each such trajectory is labeled $x_0^b$.  Each $x^b_0$ is one member
of an entire family of initial positions $x^{(b,c)}_0 = x^b_0 - c p_0
T/m$, indexed by an integer $c$; note that $x^{(b,0)}_0 = x^{b}_0$.
Thus, $b$ (branch) labels trajectories within one oscillation cycle of
Fig.~\ref{Single-Barrier-Figtommy1}(b), and $c$ (cycle) distinguishes
trajectories between different oscillation cycles.

The primitive form of the momentum-space wavefunction is given by
summing \eqref{gen} over the double index $j = (b, c)$:
\begin{align}
\tilde\Psi(p,t) =& \sum_b \sum_{c = -\infty}^\infty
F\left(x_0^{(b,c)}(p,t)\right)\left| \frac{1}{\tilde{\mathcal{J}}_{(b,c)}(p,t)}
\right|^{1/2} \nonumber \\
&\times \exp\left( \frac{i\tilde{\mathcal{S}}_{(b,c)}(p,t)}{\hbar}
- \frac{i \pi \tilde{\mu}_{(b,c)}}{2}\right).
\end{align}
Here, we allow $c$ to range over all integers, since the initial
profile $F(x_0)$ serves to effectively eliminate any trajectories that
begin outside $L$.  Since two trajectories with initial
positions $x^{(b,c)}_0$ and $x^{(b,c')}_0$, having the same $b$ index,
differ only in their (uniform) motion outside of the barrier region,
they have the same Jacobian and Maslov index, i.e.
\begin{align}
\tilde{\mathcal{J}}_{(b,c)}(p,t) &=
\tilde{\mathcal{J}}_{(b,0)}(p,t) \equiv \tilde{\mathcal{J}}_{b}(p,t), \n\\
\tilde{\mu}_{(b,c)} &= \tilde{\mu}_{(b,0)} \equiv \tilde{\mu}_b.
\label{kmr7}
\end{align}

\begin{figure}
\includegraphics[width=1\columnwidth]{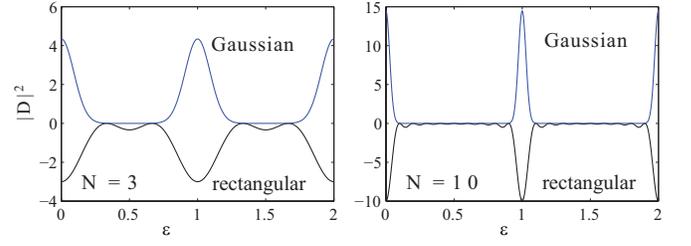}
\caption{\label{fig2} Plot of $|D|^2$ for an initial rectangular
  [lower curve, Eq.~(\ref{KMr4})] and Gaussian [upper curve,
  Eq.~(\ref{KMr6})] initial packet profiles. For the rectangular case,
  $N = 3, 10$, showing convergence to delta functions at integer
  values of $\epsilon$.  The widths $\beta$ of the Gaussian packets
  are chosen to match the standard deviations of the corresponding
  rectangular packets.}
\end{figure}
The actions too can be related to one another.  Considering first
$\tilde{\mathcal{S}}_{(b,0)}$ and $\tilde{\mathcal{S}}_{(b,1)}$, the
$(b,0)$ and $(b,1)$ trajectories follow the same path in the barrier
region, but the $(b,1)$ trajectory spends one more cycle to the left of the barrier,
whereas the $(b,0)$ trajectory spends one more cycle to the right.
Hence by Eq.~\eqref{hah2}
\begin{equation}
  \tilde{\mathcal{S}}_{(b,1)}(p,t) -
  \tilde{\mathcal{S}}_{(b,0)}(p,t) = \Delta E \, T,
\end{equation}
where $\Delta E = p^2/2 - p_0^2/2$ is the energy gained (or lost) by
the trajectory due to scattering from the barrier.  Since $\Delta E$
does not depend on the indices $b$ or $c$, we conclude that
\begin{equation}
\tilde{\mathcal{S}}_{(b,c)} = \tilde{\mathcal{S}}_{(b,0)} + c \Delta E
T \equiv  \tilde{\mathcal{S}}_b + c \Delta E T.
\label{kmr5}
\end{equation}
Eqs.~\eqref{kmr7} and ~\eqref{kmr5} provide an efficient method for constructing terms when computing the
semiclassical wavefunction.  Rather than directly integrating the
entire line of initial conditions $L$, one only needs to
integrate trajectories for initial conditions within one cycle, e.g. the interval $I$, and construct
$\tilde{\mathcal{S}}_{(b,c)}$ for other branches via \eqref{kmr5}.
The semiclassical sum can thus be rewritten as
\begin{multline}
  \tilde\Psi(p,t) = \\
  \sum_b D_b(p,t)
 \left| \frac{1}{\tilde{\mathcal{J}}_b(p,t)} \right|^{1/2}
  \exp\left( \frac{i\tilde{\mathcal{S}}_b(p,t)}{\hbar} - \frac{i \pi
      \tilde{\mu}_b}{2}\right),
\label{KMr3}
\end{multline}
where
\begin{equation}
  D_b(p,t) = \sum_{c = -\infty}^{\infty}
    F\left(x_0^b(p,t)-cp_0T/m\right) e^{ic\Delta E T/\hbar}.
\label{KMr1}
\end{equation}
In most of our calculations, we perform this sum numerically. However, in some cases, the sum can be expressed in closed form.

We consider Eq.~(\ref{KMr1}) for two initial packet profiles,
rectangular and Gaussian.  Considering the rectangular profile first,
take $F(x_0) = F_0$ constant over an interval of length $N p_0 T$,
corresponding to $N$ oscillation cycles, and $F(x_0) = 0$ outside this
interval.  Then Eq.~(\ref{KMr1}) can be rewritten as
\begin{align}
D_b(p,t)  &= D(p) = F_0 \sum_{c=0}^{N-1}  e^{2 \pi i  c \epsilon}
\nonumber \\
& = F_0 e^{2 \pi i \epsilon (N-1)/2} \frac{\sin ( \pi\epsilon N)}{\sin(\pi\epsilon)} ,
\label{KMr4}
\end{align}
where $\epsilon = \Delta E /(\hbar \omega)$.  Since $D(p)$ does not depend
on $b$, Eq.~(\ref{KMr3}) factors into the product of $D(p)$, involving
only a $c$ sum, and a quantity involving only a sum over $b$.
\begin{figure}[t]
\includegraphics[width=1\columnwidth]{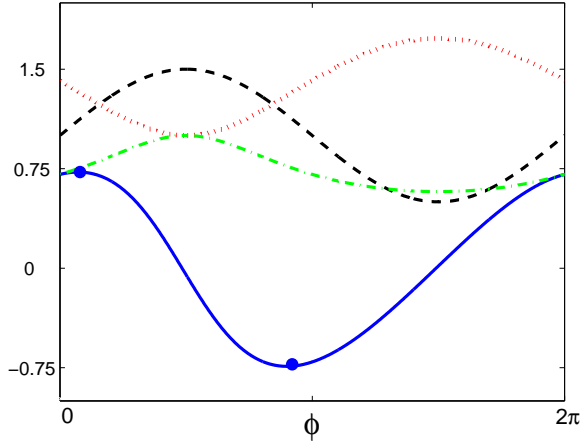}
\caption{(Color Online) For the ``elevator'' model, the change in energy vs. phase of the oscillation, $\phi$ (solid line, blue online); potential energy $V(x=0,t=0)$ (dotted line, black online); $p'$ (dash-dot line, red online), and $t_b$ (dashed line, green online). Parameters are $p_{0} = 2.0$, $U_0=1$, $A=0.5$, $m=1$, $\omega=1$, $L=0.1$. The energy change is plotted as $\Delta E/(m \omega L/p_0)$. The points on the change in energy curve, going from left to right, correspond to Eq. \eqref{deltaEhat} evaluated with our chosen parameters for $\phi=(\phi_{>}, \phi_{<})$ (see Eqs. \eqref{phihat} and \eqref{piminusphihat}), respectively.}\label{elevpic}
\end{figure}
\begin{align}
&\tilde\Psi(p,t) = \nonumber \\
& D(p)  \sum_b  \left| \frac{1}{\tilde{\mathcal{J}}_b(p,t)}
\right|^{1/2}
\exp\left( \frac{i\tilde{\mathcal{S}}_b(p,t)}{\hbar}
- \frac{i \pi \tilde{\mu}_b}{2}\right).
\label{KMr2}
\end{align}
As the length of the initial wave packet goes to infinity (i.e. $N$ goes to
infinity), $\tilde{\Psi}$ approaches a comb of delta functions
according to
\begin{equation}
  \lim_{N \rightarrow \infty} \frac{\sin ( \pi \epsilon
    N)}{\sin(\pi \epsilon)}
= \sum_{k= -\infty}^\infty  \delta(\epsilon -  k).
\end{equation}
Thus, the scattered wavefunction obeys $\Delta E = k \hbar \omega$, in
agreement with Floquet theory.  Convergence to the delta functions is
illustrated by the lower curves in Fig.~\ref{fig2}, which show $D(p)$
(Eq.~(\ref{KMr4})) as a function of $\epsilon$ for $N = 3$ and $10$.

Considering the Gaussian profile next, we now take $F(x_0)$ equal to $F_{G}\left(x_{0}\right)$
in Eq. \eqref{initgauss}.  Then in
the limit of a long packet ($\beta >> p_0 T$), Eq.~(\ref{KMr1})
reduces to
\begin{equation}
D_b(p,t)  = D(p) = \frac{1}{\sqrt{\beta} (2 \pi)^{1/4}}
\theta_3\left(\pi \epsilon, e^{-(p_0 T)^2/(2 \beta)^2}\right),
\label{KMr6}
\end{equation}
where $\theta_3(z,q)$ is a Jacobi theta function~\cite{Abramowitz65},
\begin{equation}
\theta_3(z,q) = 1 + 2\sum_{n = 1}^\infty q^{n^2} \cos(2 n z).
\end{equation}
The upper curves in Fig.~\ref{fig2} illustrate $D$ (Eq.~(\ref{KMr6}))
as a function of $\epsilon$.  As in the case of a rectangular initial
condition, $D$ converges to a comb of delta functions as the initial
packet width increases.  Unlike the rectangular case, however, there
are no higher order peaks between the primary peaks at integer values
of $\epsilon$. This agrees with the results presented in the paper (Figs.~\ref{Figure-2}-\ref{Figure-geez}, \ref{18comparison}-\ref{14comparison_t}), which also show no higher-order peaks between the primary Floquet peaks.

\section{Boundaries of Classically-Allowed Regions}\label{sec:bounds}

It would be nice to obtain some simple estimates of the maximum and minimum classically-allowed energy change. This turns out not to be as easy as we might wish. The simplest model is an ``elevator:''

\begin{figure}[t]
\includegraphics[width=1\columnwidth]{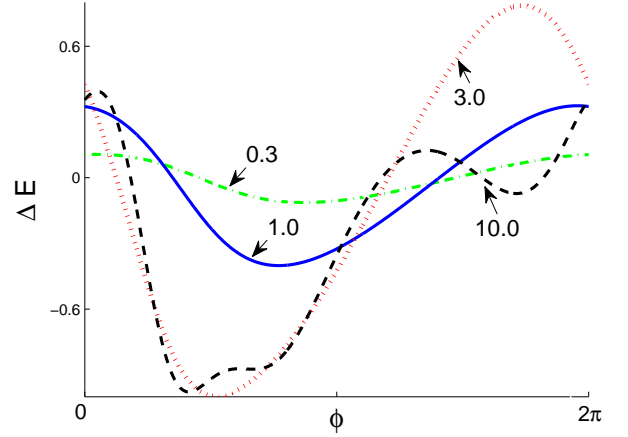}
\caption{(Color Online) Energy changes vs. phase for the same parameters as in Fig.~\ref{elevpic}, except $L=0.3$, $1.0$, $3.0$ and $10$.}\label{varylength}
\end{figure}

\begin{align}
V_{1}\left(x,t\right)=\left\{
     \begin{array}{cc}
       U_{0}\left(1 + A\sin\left(\omega t + \phi \right)\right) &, 0 \leq x \leq L\\
       0 &,\mbox{  otherwise}
     \end{array}
   \right.
\end{align}
A particle of mass $m$ and initial momentum $p_0$ arrives at $x=0$ at time $t=0$. If at that instant its kinetic energy is less than $V_{1}(0,0)$, then the particle is reflected with momentum $-p_0$. Otherwise it hops onto the elevator, traverses it with momentum $p'=\left[p_{0}^{2} - 2mV_{1}(0,0)\right]^{1/2}$, and arrives at the end of the elevator at time $t_b = mL/p'$. There it hops off, gaining potential energy $V_{1}(L,t_b)$, so the final energy and the change in energy are

\begin{subequations}
\begin{align}
E_f &= \frac{p'^{2}}{2m} + V_{1}\left(L,t_b\right)\\
\Delta E &= V_{1}\left(L,t_b\right) - V_{1}\left(0,0\right)\\
&= AU_{0}\left[\sin{\left(\omega t_b + \phi\right)} -\sin{\phi}\right]\label{deltaEhat}
\end{align}
\end{subequations}
The maximum possible range of $\Delta E$ is $\pm 2AU_{0}$. It is also important to note that $t_b$ depends on $\phi$.

Intuitively we expect that if $\omega t_b$ is small, then the particle will gain the most energy if it arrives at the barrier when the elevator is most rapidly rising, i.e., if $\phi = 0$. This is a respectable guess; however, if it arrives a bit later, then it will spend a longer time on the elevator, and thereby gain more energy. Likewise, we may expect that it will lose the most energy if it arrives when the elevator is falling most rapidly, $\phi = \pi$. However, if it arrives a bit earlier, then again it stays longer on the elevator, and so it loses more energy.

A graph of $\Delta E$ vs. $\phi$ is shown in Figs.~\ref{elevpic} and~\ref{varylength} for small $\omega t_b$. The maximum increase in energy occurs when $\phi = \phi_{>}$, where
\begin{equation}
\phi_{>}\approx \frac{mAU_{0}}{\left(p_{0}^{2} - 2mU_0\right)^{2}}\label{phihat},
\end{equation}
 and the greatest decrease occurs when $\phi = \phi_{<}$, where
\begin{align}
\phi_{<} \approx \pi - \frac{mAU_{0}}{\left(p_{0}^{2} - 2mU_0\right)^{2}}\label{piminusphihat}.
\end{align}
The change in energy predicted by these values of $\phi$ are shown in Fig.~\ref{elevpic}. For wider barriers, the behavior becomes more complex.

There are also other solvable models, such as

\begin{align}
V_{2}\left(x,t\right)=\left\{
     \begin{array}{cc}
       V_{0}\left(t\right) - |x| &, 0 \leq |x| \leq V_{0}\left(t\right)\\
       0 &,\mbox{  otherwise}
     \end{array}
   \right.
\end{align}
and
\begin{align}
V_{3}\left(x,t\right)=\left\{
     \begin{array}{cc}
       V_{0}\left(t\right) - x^{2} &, 0 \leq |x|^{1/2} \leq V_{0}\left(t\right)\\
       0 &,\mbox{  otherwise}
     \end{array}
   \right.
\end{align}
where $V_{0}\left(t\right) = U_{0}\left(1+A\sin{\left(\omega t\right)}\right)$, but they are more complicated.


\begin{thebibliography}{43}
\bibitem[{\citenamefont{it~is an~integrable singularity}()}]{sing}
\bibinfo{author}{\bibnamefont{it~is an~integrable singularity}}.
\end{thebibliography}
\end{document}